\documentclass[print]{aa}
\usepackage[latin1]{inputenc}
\usepackage{amsmath}
\usepackage{amsfonts}
\usepackage{amssymb}
\usepackage{graphicx}
\usepackage{subfig}
\usepackage{natbib}
\usepackage{hyperref}
\bibpunct{(}{)}{;}{a}{}{,}

\newcommand{\Msun}{\ensuremath{\,{\rm M}_\odot}}                  
\newcommand{\Rsun}{\ensuremath{\,{\rm R}_\odot}}                  
\newcommand{\Mjup}{\ensuremath{\,{\rm M}_{\rm Jup}}}              
\newcommand{\Rjup}{\ensuremath{\,{\rm R}_{\rm Jup}}}              
\newcommand{\pjup}{\ensuremath{\,\rho_{\rm Jup}}}                 
\newcommand{\Teff}{\ensuremath{T_{\rm eff}}}                      
\newcommand{\FeH}{\ensuremath{\left[\frac{\rm Fe}{\rm H}\right]}} 

\begin{document}

 \title{The Transiting System GJ1214: High-Precision Defocused Transit Observations\thanks{by the MiNDSTEp collaboration from the Danish 1.54m telescope at the ESO La Silla Observatory} and a Search for Evidence of Transit Timing Variation}

   \author{K.\ B.\ W.\  Harps{\o}e \inst{\ref{inst4},\ref{inst21}}\and                                 
        S.\ Hardis \inst{\ref{inst4},\ref{inst21}} \and                                       
        T.\ C.\ Hinse \inst{\ref{inst3}} \and                                     %
        U.\ G.\ J{\o}rgensen \inst{\ref{inst4},\ref{inst21}} \and     
        L.\ Mancini \inst{\ref{inst8},\ref{inst10},\ref{inst17}} \and   
        J.\ Southworth \inst{\ref{inst1}}  \and                       
        K.\ A.\ Alsubai \inst{\ref{inst19}} \and                                                  
        V.\ Bozza \inst{\ref{inst8},\ref{inst9}} \and                            
        P.\ Browne \inst{\ref{inst2}} \and                                                    
        M.\ J.\ Burgdorf \inst{\ref{inst7}} \and                                     
        S.\ Calchi Novati \inst{\ref{inst8},\ref{inst10}} \and                                
        P.\ Dodds \inst{\ref{inst2}} \and
        M.\ Dominik\thanks{Royal Society University Research Fellow} \inst{\ref{inst2}} \and
        X.-S.\ Fang \inst{\ref{inst24}} \and 
        F.\ Finet \inst{\ref{inst11}} \and                                                    
        T.\ Gerner \inst{\ref{inst16}} \and          
        S.-H.\ Gu \inst{\ref{inst24}}    \and                                     
        M.\ Hundertmark \inst{\ref{inst2},\ref{inst13}} \and                                              
        J.\ Jessen-Hansen \inst{\ref{inst23}} \and
        N.\ Kains \inst{\ref{inst2},\ref{inst14}} \and                                        
        E.\ Kerins \inst{\ref{inst15}} \and                                                   
        H. Kjeldsen \inst{\ref{inst23}} \and
        C.\ Liebig \inst{\ref{inst2}} \and                                                    
        M.\ N.\ Lund \inst{\ref{inst23}} \and
        M.\ Lundkvist  \inst{\ref{inst23}} \and
        M.\ Mathiasen \inst{\ref{inst4}} \and
        D. \ Nesvorn\'{y} \inst{\ref{inst20}} \and                                                                           
        N.\ Nikolov \inst{\ref{inst17}} \and
        M.\ T.\ Penny \inst{\ref{inst15}} \and                                                
        S.\ Proft \inst{\ref{inst16}} \and
        S.\ Rahvar \inst{\ref{inst5},\ref{inst22}} \and                                                    
        D.\ Ricci \inst{\ref{inst11}} \and                                                    
        K.\ C.\ Sahu \inst{\ref{inst18}} \and                                                     
        G.\ Scarpetta \inst{\ref{inst8},\ref{inst9},\ref{inst10}} \and                        
        S.\ Sch\"afer \inst{\ref{inst13}} \and                                                
        F.\ Sch\"onebeck \inst{\ref{inst16}} \and
        C.\ Snodgrass \inst{\ref{inst6}} \and
        J.\ Skottfelt \inst{\ref{inst4},\ref{inst21}} \and
        J.\ Surdej \inst{\ref{inst11}}    \and                                                
        J.\ Tregloan-Reed \inst{\ref{inst1}} \and
        O.\ Wertz \inst{\ref{inst11}}
        }

\institute{
      Niels Bohr Institute, University of Copenhagen, Juliane Maries vej 30, 2100 Copenhagen \O, Denmark
      \email{harpsoe@nbi.ku.dk} \label{inst4} \and
      Centre for Star and Planet Formation, Natural History Museum of Denmark, {\O}ster Voldgade 5, 1350 Copenhagen, Denmark \label{inst21} \and
      Korea Astronomy and Space Science Institute, 776 Daedeokdae-ro, Yuseong-gu, 305-348 Daejeon, Republic of Korea \label{inst3} \and
      Astrophysics Group, Keele University, Staffordshire, ST5 5BG, United Kingdom  \label{inst1} \and
      SUPA, University of St Andrews, School of Physics \& Astronomy, North Haugh, St Andrews, KY16 9SS, United Kingdom \label{inst2} \and
      Department of Physics, Sharif University of Technology, P.\,O.\,Box 11155-9161, Tehran, Iran \label{inst5} \and
      Max-Planck-Institute for Solar System Research, Max-Planck Str.\ 2, 37191 Katlenburg-Lindau, Germany \label{inst6} \and
      Qatar Foundation, Doha, Qatar \label{inst19} \and
      Dipartimento di Fisica ``E. R. Caianiello'', Universit\`a di Salerno, Via Ponte Don Melillo, 84084-Fisciano (SA), Italy \label{inst8} \and
      HE Space Operations GmbH, Flughafenallee 24, D-28199 Bremen, Germany \label{inst7} \and
      Istituto Nazionale di Fisica Nucleare, Sezione di Napoli, Napoli, Italy \label{inst9} \and
      Istituto Internazionale per gli Alti Studi Scientifici (IIASS), 84019 Vietri Sul Mare (SA), Italy \label{inst10} \and
      Institut f\"ur Astrophysik, Georg-August-Universit\"at G\"ottingen, Friedrich-Hund-Platz 1, 37077 G\"ottingen, Germany \label{inst13} \and
      European Southern Observatory, Karl-Schwarzschild-Stra{\ss}e 2, 85748 Garching bei M\"unchen, Germany \label{inst14} \and
      Institut d'Astrophysique et de G\'eophysique, Universit\'e de Li\`ege, 4000 Li\`ege, Belgium \label{inst11} \and
      Astronomisches Rechen-Institut, Zentrum f\"ur Astronomie, Universit\"at Heidelberg, M\"onchhofstra{\ss}e 12-14, 69120 Heidelberg \label{inst16} \and
      Max Planck Institute for Astronomy, K{\"o}nigstuhl 17, D-69117 Heidelberg, Germany \label{inst17} \and
      Jodrell Bank Centre for Astrophysics, University of Manchester, Oxford Road, Manchester, M13 9PL, United Kingdom \label{inst15} \and 
      Space Telescope Science Institute, 3700 San Martin Drive, Baltimore, MD 21218, USA \label{inst18} \and
      Stellar Astrophysics Centre (SAC), Department of Physics and Astronomy, Aarhus University, Ny Munkegade 120, DK-8000 Aarhus C, Denmark \label{inst23} \and 
      National Astronomical Observatories/Yunnan Observatory, Chinese Academy of Sciences, Kunming 650011, China \label{inst24} \and
      Perimeter Institute for Theoretical Physics, 31 Caroline St. N., Waterloo ON, N2L 2Y5, Canada \label{inst22} \and
      Southwest Research Institute, Department of Space Studies, 1050 Walnut St., Suite 400, Boulder, CO 80302, USA \label{inst20}
      }

  \abstract
   {}
   {We present 11 high-precision photometric transit observations of the transiting super-Earth planet GJ\,1214\,b. Combining these data with observations from other authors, we investigate the ephemeris for possible signs of transit timing variations (TTVs) using a Bayesian approach.}
   {The observations were obtained using telescope-defocusing techniques, and achieve a high precision with random errors in the photometry as low as 1\,mmag per point. To investigate the possibility of TTVs in the light curve, we calculate the overall probability of a TTV signal using Bayesian methods.}
   {The observations are used to determine the photometric parameters and the physical properties of the GJ\,1214 system. Our results are in good agreement with published values. Individual times of mid-transit are measured with uncertainties as low as 10\,s, allowing us to reduce the uncertainty in the orbital period by a factor of two.}
   {A Bayesian analysis reveals that it is highly improbable that the observed transit times show a TTV, when compared with the simpler alternative of a linear ephemeris.}

   \titlerunning{The Transiting System GJ1214}
   \authorrunning{Kennet B. W. Harps\o e  et al.}
   \maketitle

   
\section{Introduction}

The transiting exoplanet GJ\,1214\,b was discovered in 2009 by the MEarth project\footnote{\url{https://www.cfa.harvard.edu/~zberta/mearth/Welcome.html}} \citep{mearth}. This planet transits a nearby M dwarf \citep{Charbonneau09}, with a mass $0.15$\,M$_{\sun}$ and the planet is generally classified as a super-Earth with a mass and radius ($M_{\rm p} = 6.37$\,M$_{\oplus}$ and $R_{\rm p} = 2.74$\,R$_{\oplus}$ according to  \cite{Kundurthy11}, in this study we find $M_{\rm p} = 6.26$\,M$_{\oplus}$ and $R_{\rm p} = 2.85$\,R$_{\oplus}$) between that of Earth and Neptune, a type of planet that has no solar system analogue. GJ\,1214\,b is one of the { lowest temperature} transiting exoplanets known and, as it is also detectable with radial velocity (RV) methods -- a very interesting target for detailed study.

Due to the relatively low mean density of GJ\,1214\,b, $\rho_{p} = 1.49 \pm 0.33$\,g\,cm$^{-3}$ in this study, it has been suggested to hold some extended atmosphere or gaseous envelope. But the planet composition in this mass and radius range is degenerate \citep{Adams08}, warranting further studies in order to determine its composition. Defining what the atmosphere consists of can help determine the planet composition. \cite{RogersSeager10} describes three possibilities for the interior and atmospheric composition of GJ\,1214\,b. It could be (1) a mini-Neptune with a H/He gas envelope, (2) a ``water world'' with a water-rich and ice-dominated interior and a water-vapour-dominated envelope, or (3) a rocky planet with an atmosphere mainly consisting of H$_2$.

Recent results from among others the {\it Kepler} mission \citep{2011ApJ...732L..24L} and gravitational microlensing \citep{2010ApJ...720.1073G} gives reason to believe that multiple systems are common. It is therefore inherently interesting to look for traces of transit timing variations in any transiting system and especially so in GJ\,1214, as the planet seems to be at the inner edge of the habitable zone, with an equilibrium temperature of in the region of 393 to 555\,K \citep{Charbonneau09}, i.e. finding a planet in a slightly larger orbit would be very interesting.  

Additional planets can be revealed via their gravitational effects on the transiting planet. This would result in telltale systematic deviations in the mid-transit times from a linear ephemeris, a phenomenon known as transit timing variation (TTV) \citep{Agol05,2005Sci...307.1288H}.  GJ\,1214\,b is well-suited to such an analysis due to its short transits, allowing precise measurements of mid-transit times, and its short period, which means many transits are observable. One disadvantage is its relative faintness, which could cause a loss of precision in the measured transit times.

We have gathered photometric observations of 11 transits in the GJ\,1214 system, and modelled them to estimate the transit times. Inclusion of results from the literature leads to a total time interval of 833 days over which TTVs can be investigated. In the following work we cast the problem of detecting a TTV signal as a model selection problem and via Bayesian methods calculate the probability that the data, as a whole, actually contain a TTV signal. 

In section 2 we present the new data and their reduction, which in section 3 are analysed in order to obtain new transit times and physical properties. Section 4 contains a description of the Bayesian model selection process.

\begin{table*}
\caption{Log of the transit observations of GJ\,1214 for this work.}
\label{tab:obslog} \centering
\begin{tabular}{llccccccccc}
\hline\hline
Date & Telescope  & Start & End   & Number of & Exposure & Filter & Airmass & Scatter & Aperture   & PSF area \\
     & instrument & (UTC) & (UTC) & exposures & time (s) &        &         & mmag    & sizes\tablefootmark{a}
 (px) & (px$^2$) \\
\hline
2010/07/06 & DFOSC & $05\mbox{:}01$ & $06\mbox{:}44$ & $33$  & $150$         & $I$ & $1.40\mbox{--}2.18$ & $0.97$ & 21,57,77   & $1385$ \\
2010/07/14 & DFOSC & $00\mbox{:}56$ & $05\mbox{:}59$ & $142$ &$80\mbox{--}90$& $I$ & $1.33\mbox{--}2.02$ & $1.18$ & 19,30,46   & $1134$ \\
2010/08/02 & DFOSC & $00\mbox{:}25$ & $04\mbox{:}31$ & $154$ & $60$          & $I$ & $1.24\mbox{--}1.87$ & $0.99$ & 15.5,28,46 & $755$  \\
2010/08/21 & DFOSC & $00\mbox{:}02$ & $03\mbox{:}58$ & $154$ & $60$          & $I$ & $1.21\mbox{--}2.45$ & $1.29$ & 16,26,48   & $804$  \\
2011/06/03 & DFOSC & $02\mbox{:}33$ & $05\mbox{:}11$ & $101$ & $60$          & $R$ & $1.61\mbox{--}1.21$ & $1.36$ & 16,24,45   & $804$  \\
2010/08/06 & BFOSC & $19\mbox{:}52$ & $22\mbox{:}55$ & $71$  & $90$          & $i$ & $1.30\mbox{--}1.61$ & $1.74$ & 12,20,40   & $452$  \\
2011/08/25 & BUSCA & $19\mbox{:}52$ & $22\mbox{:}58$ & $123$ & $90$          & $g$ & $1.19\mbox{--}1.30$ & $3.20$ & 8,16,24    & $201$  \\
2011/08/25 & BUSCA & $19\mbox{:}52$ & $22\mbox{:}58$ & $125$ & $90$          & $r$ & $1.19\mbox{--}1.30$ & $1.92$ & 20,55,65   & $1257$ \\
2011/08/25 & BUSCA & $19\mbox{:}52$ & $22\mbox{:}58$ & $125$ & $90$          & $z$ & $1.19\mbox{--}1.30$ & $2.66$ & 25,35,45   & $1963$ \\
2010/04/29 & GROND & $05\mbox{:}52$ & $08\mbox{:}38$ & $89$  & $50$          & $i$ & $1.31\mbox{--}2.54$ & $3.65$ & no defocus & $--$   \\
2010/04/29 & GROND & $05\mbox{:}52$ & $08\mbox{:}38$ & $90$  & $50$          & $r$ & $1.31\mbox{--}2.54$ & $4.80$ & no defocus & $--$   \\

\hline
\end{tabular}
\tablefoot{
\tablefoottext{a}{The three numbers are the apertures radii in pixels of the object aperture and the inner and outer edge of the sky annulus.}}

\end{table*}

   
\section{Observations and Data Reduction}

We observed five transits of GJ\,1214\,b in the period of 2010 July 6th to 2011 June 3rd, using the Danish 1.54\,m telescope at ESO La Silla and the focal-reducing camera DFOSC. The plate scale of DFOSC is $0.39\arcsec$ per pixel. The full field of view is $13.7\arcmin \times 13.7\arcmin$, but for each transit observation the CCD was windowed down to reduce the readout time from around 90\,s to approximately 30\,s. The four transits in 2010 were observed through a Cousins \textit{I} filter, while the 2011 transit was observed in Johnson \textit{R}. An observing log is given in Table\,\ref{tab:obslog}.  

The transits were observed with the telescope defocused, in order to use longer exposure times whilst avoiding CCD saturation. This approach allowed us to decrease the Poisson and scintillation noise by exposing for a larger fraction of the time during transit (see \citealt{Southworth09a}). The impact of flat-fielding errors was minimised by the use of defocusing and by autoguiding the telescope throughout the observations. The diameters of the defocused point spread functions (PSFs) ranged from 31 to 42 pixels.

We also observed one transit of GJ\,1214\,b using the 1.52\,m Cassini Telescope at the Loiano Observatory, Italy. The BFOSC CCD imager was used, and was defocused with the same approach as taken for the Danish telescope observations (see \citealt{2010MNRAS.408.1680S}).

One transit was observed simultaneously in the $g$, $r$ and $z$ filters using the Calar Alto 2.2\,m telescope and the BUSCA four-beam CCD imager. A fourth dataset was obtained in the $u$-band but, as expected, yielded a light curve which was too noisy to be useful. For further discussion on the observing strategy we used for BUSCA please see \citet{2012MNRAS.422.3099S}.

Finally, one transit was observed in 2010 using the GROND seven-beam imager on the 2.2\,m MPI telescope at ESO La Silla. The observations were performed without telescope defocusing. Useful results could be obtained for only the $r$ and $i$ filters, with the star proving too faint in the $g$, $z$ and $JHK$ channels.

The data were reduced using the pipeline described in \cite{Southworth09a}, which performs standard aperture photometry with the \textsc{astrolib/aper}\footnote{Distributed within the {\sc idl} Astronomy User's Library at \url{http://idlastro.gsfc.nasa.gov/}} \textsc{idl} routine. A range of aperture sizes were tried, and the ones which gave the least noisy light curves were adopted (Table\,\ref{tab:obslog}). The form of the transit is insensitive to the choice of aperture sizes, and to the presence of two faint stars within the sky annulus.

Comparison stars to be used for relative photometry were chosen within the field, and the \textsc{astrolib/aper} routine determined relative magnitudes for these and GJ\,1214, from the given coordinates and aperture radius. Potential comparison stars that proved to be variable or too faint were discarded. Relative photometry of GJ\,1214 was obtained against an optimally weighted ensemble of comparison stars.

\begin{figure}
\includegraphics[width=1.0\linewidth]{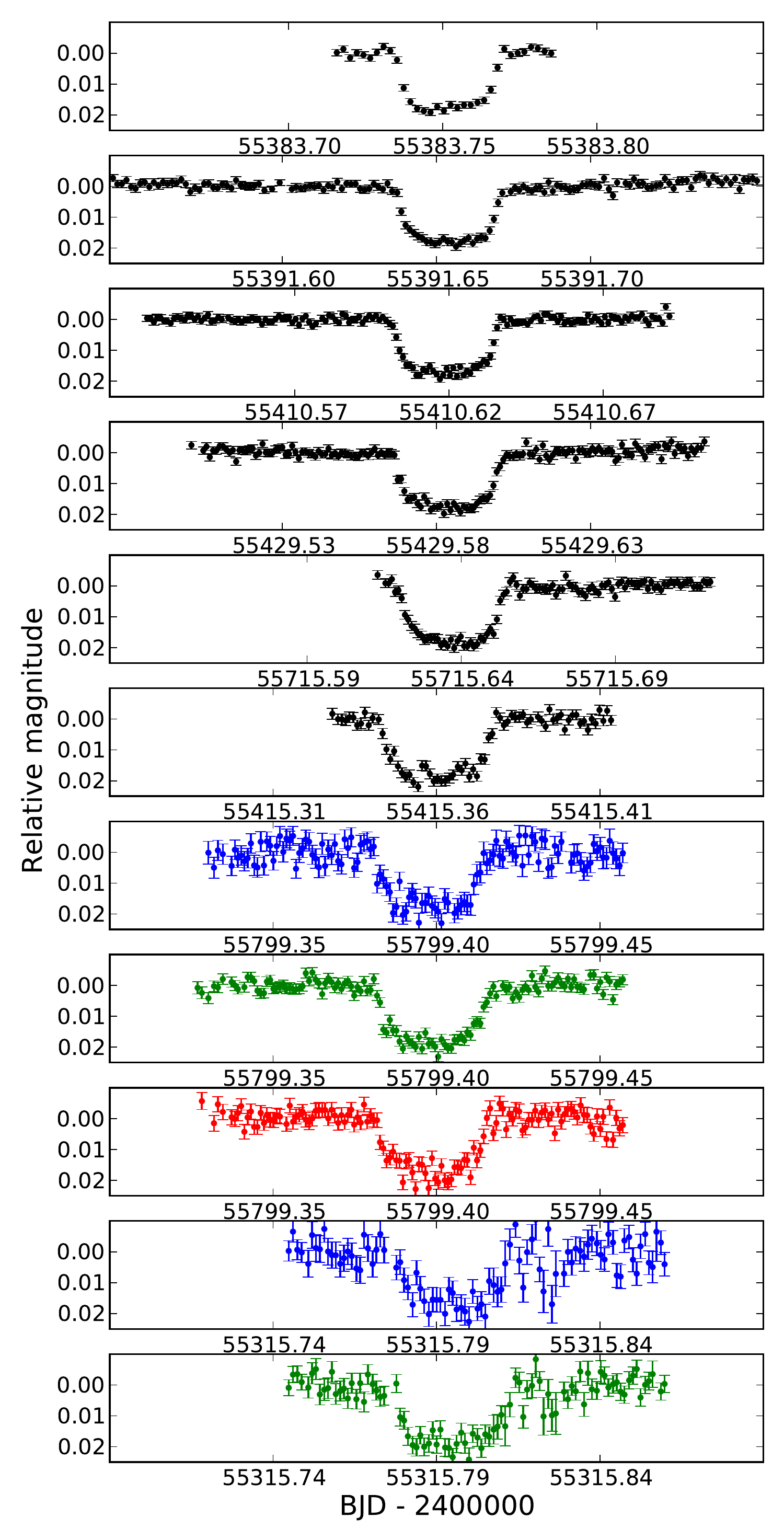}
\caption{The 11 observed transit light curves of GJ\,1214\,b, plotted in the same order as listed in the observing log in Table \ref{tab:obslog}. For the observations taken in multiple filters simultaneously the datapoints are coloured in spectral order, i.e.\ blue for bluest etc.}
\label{fig:nofitlc}
\end{figure}

The resulting GJ\,1214 light curve does not have a constant magnitude out of transit, primarily due to changes in airmass and intrinsic stellar variability. To correct any systematic trends, the out-of-transit data points were fitted with a straight line. Simultaneous optimisation of the comparison star weights and the out-of-transit polynomial was used to obtain the final light curves, which are shown in Fig.\,\ref{fig:nofitlc}.

Three additional light curve were obtained from the Exoplanet Transit Database\footnote{\url{http://var2.astro.cz/ETD/index.php}} \citep{etd}. These were contributed by Johannes Ohlert (2010/07/18) and Thomas Sauer (2010/06/29 and 2010/07/07).

\section{Light curve analysis}

\subsection{Analysis with \textsc{jktebop}}

The light curves were analysed with the \textsc{jktebop}\footnote{\textsc{jktebop} is written in \textsc{fortran77} and the source code is available at \url{http://www.astro.keele.ac.uk/~jkt/}} code \citep{Southworth04a,2004MNRAS.351.1277S}, originally developed as {\sc ebop} \citep{PopperEtzel81,Etzel81} for modelling light curves of detached eclipsing binaries. The use of the code for exoplanet transit light curves is discussed in detail in \cite{Southworth08}.

The size of the planet relative to the size of the star is directly related to the transit depth. \textsc{jktebop} models the sky projections of the two objects as biaxial spheroids, dividing them into concentric circles and assigning a limb darkening to each of the rings before estimating the flux. With optical observations of a planetary system, it is safe to assume that the secondary object, the exoplanet, is dark, so the surface brightness ratio can thus be set to zero. Using \textsc{jktebop} we fitted for the inclination $i$ of the orbit, the sum and ratio of the fractional radii $k = r_{\rm p} + r_{\star}$ and $r_{\rm p}/r_{\star}$. The fractional radii are $r_{\rm p} = R_{\rm p}/a$ and $r_{\star} = R_{\star}/a$, where $R_{\rm p}$ and $R_{\star}$ are the absolute planetary and stellar radii, respectively, and $a$ is the semi-major axis of the orbit. We also fitted for the time of minimum of each light curve, $T_{mid}$, using a fixed orbital period of $P=1.58040490$ days \citep{Berta11}. The mass ratio, which only affects the shape of the ellipsoids describing the components, was fixed at $q=0.0002$, {\rm which is sufficiently close to the value 0.00013 found in this paper. We found that changes, less than one order of magnitude, in this value have a negligible effect on the results.}

Limb darkening (LD) affects the shape of a transit light curve. LD will cause the bottom of the transit to have a curved shape. We tried fitting four different LD laws: linear, square-root, logarithmic and quadratic. LD coefficients can be found from stellar atmosphere models given the effective temperature and surface gravity. Only \cite{Claret00,claret2004} provides LDCs for stars as cool as GJ\,1214\,A. We used values for a star of $T_{\rm eff} = 3000$\,K and $\log g = 5.00$ (cgs units).

\begin{figure}
\resizebox{\hsize}{!}{
\includegraphics[width=1.0\linewidth]{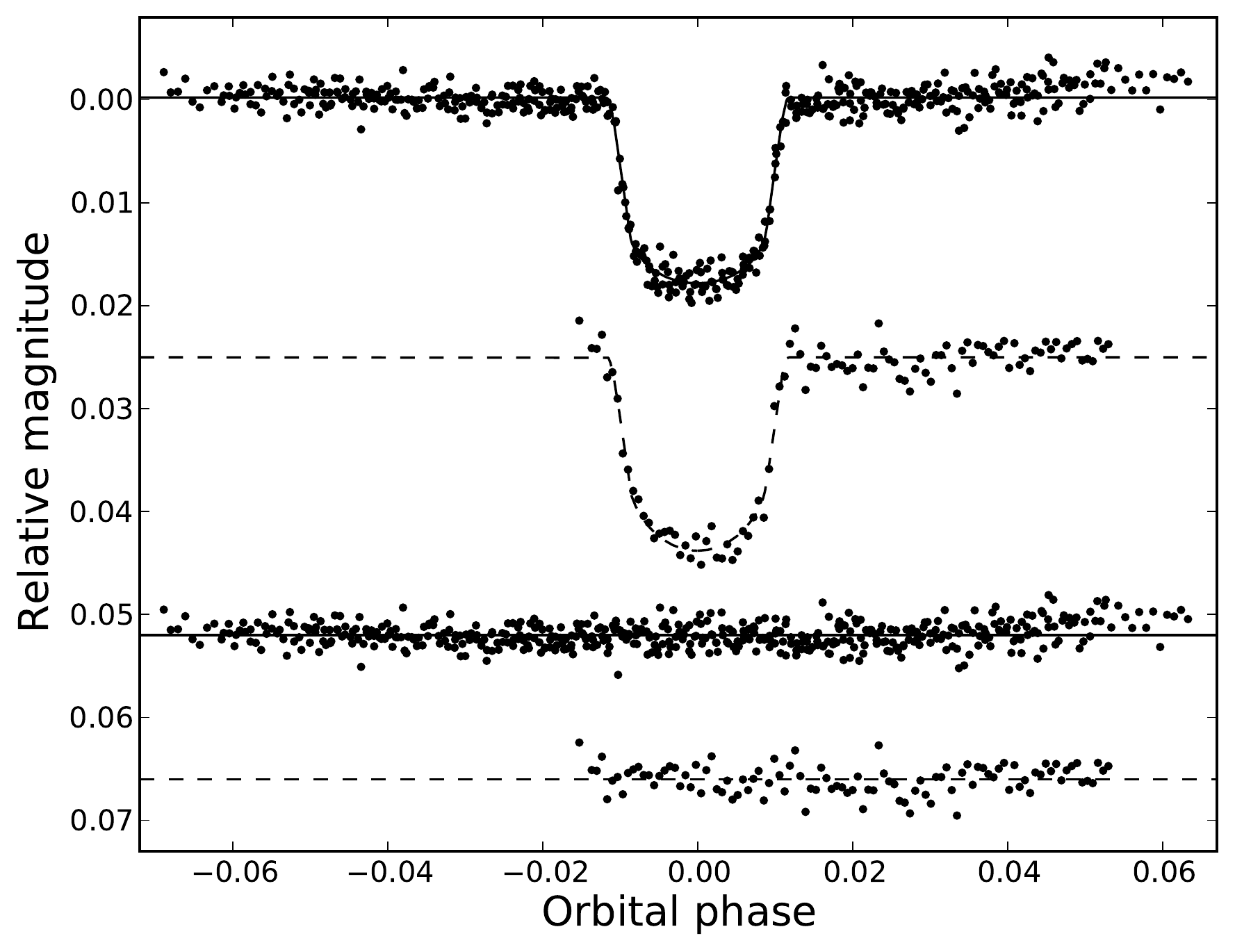}}
\caption{Plot of the combined Danish telescope light curve of the 2010 data (top) and the 2011 light curve bottom), 
versus the best \textsc{jktebop} fit (solid lines 2010, dashed 2011) with the logarithmic LD law. 
The residuals to the fits are plotted below, where the lines mark zero residual.}
\label{fig:finalplot}
\end{figure}

We analysed the combined 2010 Danish telescope data, finding that one LD coefficient could be included as a fitted parameter. We therefore fitted for the linear coefficient whilst holding the nonlinear coefficient fixed at the theoretical value. For the other datasets we had to fix both coefficients in order to avoid unphysical results. The best fits to the Danish telescope data are plotted in Fig.\,\ref{fig:finalplot}. 

In order to obtain uncertainties in the fitted parameters we first rescaled the error bars of the datapoints to give a reduced $\chi^2$ of unity for each transit light curve. This step is necessary because the data errors returned by the {\sc aper} algorithm are normally too small. We then performed 1000 Monte Carlo simulations (see \citealt{2004MNRAS.351.1277S}) on each light curve to derive the errorbars in the fitted parameters quoted in Table\,\ref{tab:parameters1}.

The light curve from 2010/07/06 was unintentionally obtained using an exposure time of 150\,s, which is significant compared to the duration of the ingress and egress. When fitting these data we used the possibility within \textsc{jktebop} to numerically integrate the light curve model in order to obtain an unbiased fit \citep{Southworth11numint}.

\begin{table*}
\caption{Photometric parameters of GJ\,1214 from the best-fitting light curves to the 2010-season 
data from the Danish telescope. $\sigma$ is the rms scatter of the data around the best fit.}
\label{tab:parameters1}
\centering
\begin{tabular}{lcccc}
\hline\hline
LD law                  & Linear                 & Square-root              & Logarithmic             & Quadratic               \\
\hline
$r_{\star}+r_{p}$       & $0.0812 \pm 0.0030$     & $0.0803 \pm 0.0032$     & $0.0805 \pm 0.0033$     & $0.0777 \pm 0.0040$     \\
$k$                     & $0.1222 \pm 0.0014$     & $0.1209	\pm 0.0015$     & $0.1212 \pm 0.0016$     & $0.1190 \pm 0.0021$     \\
$i$                     & $87.97^{+0.37}_{-0.28}$ & $88.12^{+0.42}_{-0.33}$ & $88.09^{+0.42}_{-0.35}$ & $88.50^{+0.78}_{-0.50}$ \\
$u$                     & $0.45 \pm 0.06$         & $0.05 \pm 0.06$         & $0.59 \pm 0.06$         & $0.28 \pm 0.07$         \\
$v$                     &                         & $0.70$ fixed            & $0.20$ fixed            & $0.40$ fixed            \\
$r_{\star}$             & $0.0724 \pm 0.0027$     & $0.0717 \pm 0.0028$     & $0.0718 \pm 0.0028$     & $0.0694 \pm 0.0035$     \\
$r_{p}$                 & $0.00886 \pm 0.00039$   & $0.00866 \pm 0.00042$   & $0.00870 \pm 0.00044$   & $0.00826 \pm 0.00055$   \\
$\sigma$ (mmag)         & $1.18$                  & $1.18$                  & $1.18$                  & $1.18$                  \\

\hline
\end{tabular}
\end{table*}

\subsection{Physical properties of the system}

It is possible to calculate the physical properties of the GJ\,1214 system from our measured photometric parameters and from published values of the stellar effective temperature, metal abundance and orbital velocity amplitude. We obtained final values for $r_\star$, $r_{\rm p}$ and $i$ from our results for the 2010 Danish Telescope data. Each was calculated as the mean of the values from the solutions using the four LD laws, with uncertainty the quadrature sum of the largest individual uncertainty plus the standard deviation of the four individual parameter values. For the star we adopted the temperature $\Teff = 3026 \pm 150$\,K and orbital velocity amplitude $K_\star = 12.2 \pm 1.6$\,m\,s$^{-1}$ from \citet{Charbonneau09}, and the metal abundance $\FeH = +0.39 \pm 0.15$ from \citet{2010ApJ...720L.113R}.

The physical properties were then calculated by requiring the properties of the star to match the tabulated predictions of the DSEP stellar evolutionary models \citep{2008ApJS..178...89D}. This step was performed using the procedure outlined by \citet{2009MNRAS.394..272S}. The DSEP model set was chosen because it is the only one of the five sets used by \citet{2010MNRAS.408.1689S} which reaches to sufficiently low stellar masses. 

The input and output parameters for this analysis are given in Table\,\ref{tab:absdim} and show a reasonable agreement with literature values. The mass, radius and surface gravity of the star are given by $M_{\star}$, $R_{\star}$ and $\log g_{\star}$, respectively. The mass, radius, surface gravity and mean density of the planet are denoted by $M_{\rm p}$, $R_{\rm p}$, $g_{\rm p}$ and $\rho_{\rm p}$, respectively.

{ The measured physical properties will be subject to systematic errors as theoretical evolutionary models are not perfect representations of reality. \citet{2009MNRAS.394..272S} found that this systematic error was generally 1\% or less for the masses and radii of transiting planets and their host stars. In the case of GJ 1214 this systematic error could be significantly larger, due to the relatively poorer theoretical understanding of 0.2 \Msun stars, but will still be significantly smaller than the statistical uncertainties quoted in Table \ref{tab:absdim}.}

\begin{table} \centering \caption{\label{tab:absdim} Derived physical properties 
of the GJ\,1214 system. The upper part of the table contains the input parameters 
to the property-calculation algorithm and the lower part the output parameters.}
\begin{tabular}{l r@{\,$\pm$\,}l}
\hline \hline
$r_\star$                     & 0.0713  & 0.0037  \\
$r_{\rm p}$                   & 0.00862 & 0.00059 \\
$i$ (deg)                     & 88.17   & 0.54    \\
\Teff\ (K)                    & 3026    & 150     \\
\FeH\ (dex)                   & +0.39   & 0.15    \\
$K_\star$ (m\,s$^{-1}$)       & 12.2    & 1.6     \\ 
\hline
$M_{\star}$           (\Msun) & 0.150   & 0.011   \\
$R_{\star}$           (\Rsun) & 0.216   & 0.012   \\
$\log g_{\star}$        (cgs) & 4.944   & 0.013   \\
$M_{\rm p}$           (\Mjup) & 0.0197  & 0.0027  \\
$R_{\rm p}$           (\Rjup) & 0.254   & 0.018   \\
$g_{\rm p}$     (m\,s$^{-2}$) & 7.6     & 1.5     \\
$\rho_{\rm p}$        (\pjup) &  1.12   & 0.25    \\
$a$                      (AU) & 0.01411 & 0.00032 \\
\hline \end{tabular} \end{table}

\subsection{Orbital period determination}
\label{sec:period}
The times are given in Barycentric Julian days (BJD), and have been calculated from UTC with codes provided by \cite{Eastman2010}. By augmenting our measured $T_{mid}$ values with ones from the literature (Table \ref{tab:timemin}), we are able to refine the orbital ephemeris for GJ\,1214, refitting for $P$ and $T_0$ the zero epoch. Taking the second observed transit to be the reference epoch we find the ephemeris: 
\begin{align}
T_{mid} & = {\rm BJD(TDB)}\ 2455320.535733 \pm 2.1\cdot 10^{-5} \\
                & + 1.58040456 \pm 1.6\cdot10^{-7} \times E \nonumber
\end{align}
where $E$ is the orbit count with respect to the reference epoch. The reduced $\chi^2$ for this fit is 1.24. The estimated period is identical to the previous estimates, but the uncertainty on the period $P$ has been reduced by approximately a factor of two \citep{Berta11,Carter11,Kundurthy11}. The fit and the residuals are plotted in Fig.\,\ref{fig:resi}. Given the rather large spread in data points compared with the period, one would not expect this estimate of the period to be afflicted by significant systematic error. On the other hand the $T_0$ could have systematic error, which is handled in the later Bayesian analysis by marginalising out this parameter. 


\section{Transit timing variation as Bayesian model selection}

\begin{figure*}
\includegraphics[width=0.7\textwidth,angle=-90]{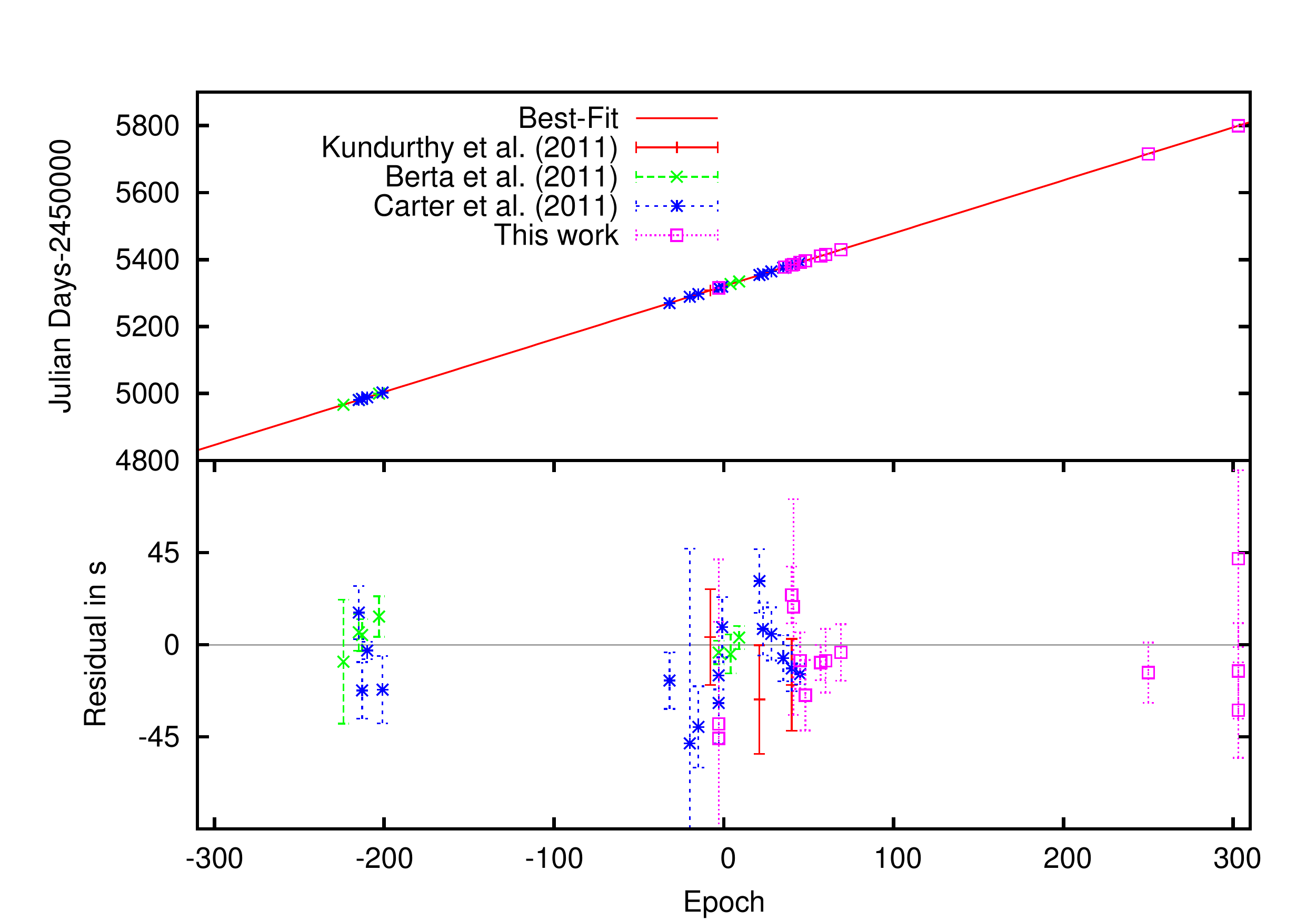}
\caption{Times of mid-transit from the literature and this work. In the upper plot the errorbars are smaller than the point marker and are thus not shown. The lower plot shows the residuals on a different scale. Some of the observations are taken with instruments that simultaneously observe in several filters so some of the data points coincide in time. }
\label{fig:resi}
\end{figure*}

A system with a transiting planet enables the possibility of detecting additional unseen planets in the timing data by TTVs, that is, look for systematic trends in the residuals of Fig.\,\ref{fig:resi}. We will in the following outline a method for quantitatively assessing the probability of whether such a signal exists in the timing dataset or not.

One could simply { fit an appropriate function describing mutual gravitational perturbations}, and evaluate a measure like the classical squared sum of residuals. But it is, in general, true that a model with more free parameters { will be able to fit noise better, i.e. produce a lower squared sum of residuals by fitting nonphysical features}. This is the concept of over-fitting. The problem is especially prominent when the sought effect is on the same order as the uncertainty in the data, in which case it is not easy to determine how much of an improvement in the squared sum of residuals one should demand for a more complex model to be plausible.

One is forced to penalise models for having too many parameters. Thus the problem is no longer a problem of parameter estimation, but a problem of model selection. One needs to apply Occam's razor. One formal way of doing so is via Bayesian model selection, which is completely analogous to, but distinct from, Bayesian parameter estimation. Whilst it might not be possible to estimate the parameters of a model with any certainty, a wide range of plausible parameters which do improve the squared residuals by some amount would lend credibility to the notion that the model in question is true. How true can be expressed as a probability.

The search for a TTV signal can conveniently be cast as a model selection problem. If there is a TTV signal there has to be some kind of pattern in the residuals listed in Table\,\ref{tab:timemin}. The expected signal from perturbation from another body can be calculated with an N-body orbital mechanics code.
\begin{table}[]
\caption{Mid-transit times from the literature and the present work. Reference 1: This work, 2: \citet{Kundurthy11}, 3: \citet{Berta11} and 4: \citet{Carter11}.}
\centering
	\begin{tabular}{lrrc}
	\hline\hline
	$T_{mid}$	& Epoch		& Residual (O-C) & Ref.	\\
	\hline                       
	BJD(TDB)  &          &       $\times 10^{-4}$ &  \\
	\hline

$ 2454980.748682 \pm 0.000104$ & -215 & -0.702161 & 3\\ 
$ 2454980.748570 \pm 0.000150$ & -215 & -1.822165 & 4\\ 
$ 2454983.909820 \pm 0.000160$ & -213 & 2.586553 & 4\\ 
$ 2454983.909507 \pm 0.000090$ & -213 & -0.543450 & 3\\ 
$ 2454988.650808 \pm 0.000049$ & -210 & 0.329618 & 4\\ 
$ 2454999.713448 \pm 0.000115$ & -203 & -1.589879 & 3\\ 
$ 2455002.874670 \pm 0.000190$ & -201 & 2.538827 & 4\\ 
$ 2455269.962990 \pm 0.000160$ & -32 & 2.025110 & 4\\ 
$ 2455288.928200 \pm 0.001100$ & -20 & 5.577393 & 4\\ 
$ 2455296.830130 \pm 0.000230$ & -15 & 4.649176 & 4\\ 
$ 2455307.892454 \pm 0.000271$ & -8 & -0.430327 & 2\\ 
$ 2455315.794693 \pm 0.000080$ & -3 & 1.731454 & 4\\ 
$ 2455315.794850 \pm 0.000230$ & -3 & 3.301455 & 4\\ 
$ 2455315.794968 \pm 0.000930$ & -3 & 4.481454 & 1\\ 
$ 2455315.795050 \pm 0.000660$ & -3 & 5.301451 & 1\\ 
$ 2455315.794564 \pm 0.000066$ & -3 & 0.441452 & 3\\ 
$ 2455318.955230 \pm 0.000170$ & -1 & -0.989833 & 4\\ 
$ 2455326.857404 \pm 0.000110$ & 4 & 0.521950 & 3\\ 
$ 2455334.759334 \pm 0.000066$ & 9 & -0.406266 & 3\\ 
$ 2455353.724539 \pm 0.000307$ & 21 & 3.096014 & 2\\ 
$ 2455353.723870 \pm 0.000180$ & 21 & -3.593988 & 4\\ 
$ 2455356.884950 \pm 0.000150$ & 23 & -0.885273 & 4\\ 
$ 2455364.787000 \pm 0.000150$ & 28 & -0.613490 & 4\\ 
$ 2455375.849970 \pm 0.000130$ & 35 & 0.767009 & 4\\ 
$ 2455377.431461 \pm 0.000420$ & 36 & 11.631362 & 1\\ 
$ 2455383.752050 \pm 0.000130$ & 40 & 1.338790 & 4\\ 
$ 2455383.751635 \pm 0.000160$ & 40 & -2.811207 & 1\\ 
$ 2455383.752143 \pm 0.000260$ & 40 & 2.268790 & 2\\ 
$ 2455385.332107 \pm 0.000610$ & 41 & -2.136850 & 1\\ 
$ 2455391.654105 \pm 0.000059$ & 45 & 1.660576 & 4\\ 
$ 2455391.654029 \pm 0.000160$ & 45 & 0.900575 & 1\\ 
$ 2455396.395438 \pm 0.000200$ & 48 & 2.853647 & 1\\ 
$ 2455410.618895 \pm 0.000100$ & 57 & 1.012855 & 1\\ 
$ 2455415.360098 \pm 0.000180$ & 60 & 0.905925 & 1\\ 
$ 2455429.583692 \pm 0.000160$ & 69 & 0.435133 & 1\\ 
$ 2455715.637033 \pm 0.000170$ & 250 & 1.583700 & 1\\ 
$ 2455799.397830 \pm 0.000500$ & 303 & -4.865397 & 1\\ 
$ 2455799.398465 \pm 0.000270$ & 303 & 1.484603 & 1\\ 
$ 2455799.398686 \pm 0.000270$ & 303 & 3.694603 & 1\\ 

    \hline
    \end{tabular}
    \label{tab:timemin}

\end{table}

\subsection{Bayesian estimation}

\subsubsection{Parameter estimation}

Following the development in \cite[Chapter 3][]{gregory05}, Bayesian parameter estimation can conceptually be thought of as testing a range of mutually excluding hypotheses $H_i$, i.e. one assumes a parametrised model $M$. This model can be thought of as a logical disjunction (``or'') $M = H_1 + H_2 + \cdots$  where the hypothesis $H_i$ implies that a given parameter $\theta$ has the particular value $\theta_i$.

For mutually exclusive probabilities the sum rule applies. Assuming that the parameter $\theta$ does indeed take a value, one can write
\begin{equation}
\sum_i p(H_i \vert M) = 1
\end{equation}

From Bayes' theorem one learns that
\begin{equation}
p(H_i \vert D, M) = \frac{p(H_i \vert M) p(D \vert H_i, M)}{p(D\vert M)}
\end{equation}
where the left hand side $p(H_i \vert D, M)$ is the posterior probability of $H_i$, i.e. the probability of the hypothesis $H_i$ in the light of the data $D$. The term $p(D \vert H_i, M)$ is the probability of the data $D$ if $H_i$ true. This quantity is known as the likelihood of $H_i$. The term $p(H_i \vert M)$ is known as the prior and represents the probability one assigns to $H_i$ in the light of one's model before any data becomes available. Finally the term in the denominator $p(D \vert M)$ is known as the global likelihood or the evidence.

Based on assumptions one can deduce the value of $p(D \vert M)$
\begin{equation}
\sum_i p(H_i \vert D, M) = \frac{\sum_i p(H_i \vert M) p(D \vert H_i, M)}{p(D \vert M)} = 1
\end{equation}
Thus
\begin{equation}
p(D \vert M) = \sum_i p(H_i \vert M) p(D \vert H_i, M)
\label{evidence}
\end{equation}
That is, the denominator, which does not depend on the individual hypotheses $H_i$, is the sum of the numerator over $H_i$. The process of summing over all the hypothesis is known as marginalisation. It is clear that the evidence serves as a normalisation constant. For parameter estimation this normalisation is unimportant, as in most cases one simply seeks the maximum posterior value without regard to normalisation. But for model selection this evidence term is important.

\subsubsection{Model selection}

Model selection is carried out following the exact same procedure as above, only assuming a disjunction of models $I = M_1 + M_2 + \cdots + M_N$, instead of a disjunction of hypotheses.

Bayes' theorem now reads
\begin{equation}
p(M_i \vert D, I) = \frac{p(M \vert I) p(D \vert M_i, I)}{p(D \vert I)}
\label{modelselec}
\end{equation}
One immediately recognises the second term in the numerator as the evidence term from Eq.\,\ref{evidence}. Just as the probability of a parameter is proportional to the likelihood times the prior, the probability of a model is proportional to the evidence times the prior. The denominator in Eq.\,\ref{modelselec} can be calculated like the evidence in Eq.\,\ref{evidence}, assuming that one of the $M_i$ is the true model. In other words the probability of a model is given by marginalising over all the parameters in the model.

Often it is more convenient to work with odds ratios $O$ and Bayes factors $B$ defined as
\begin{equation}
O_{ij} = \frac{p(M_i \vert I)}{p(M_j \vert I)}\frac{p(D\vert M_i,I)}{p(D \vert M_j, I)} = \frac{p(M_i \vert I)}{p(M_j \vert I)} B_{ij}
\end{equation}
Assuming that
\begin{equation}
\sum_{N} p(M_i \vert D, I) = 1
\end{equation}
one can write the probabilities of the individual models as
\begin{equation}
p(M_i \vert D, I) = \frac{O_{i1}}{\sum_{N} O_{j1}}
\label{modelprob}
\end{equation}
It is crucial to notice that these odds ratios implicitly result in an Occam's razor. One may notice that the evidence term in Eq.\,\ref{evidence} takes the form of an average of the likelihood over the prior. Hence penalising complicated models by their unused prior space, e.g.\ large swathes of prior space with negligible likelihood, will pull down the average of the likelihood over the prior. In other words, the more plausible model is the one that makes more sharp predictions with fewer adjustable parameters. Note that the form of the priors and the prior ranges are as much part of the model specification as the functional form of the likelihood function, which is no surprise given that probabilities are purely a function of one's state of knowledge. There is nothing inherent in nature about probabilities. Also note that assigning probabilities according to Eq.\,\ref{modelprob}, i.e. normalising to a total sum of one, implicitly assumes that one of the models is true.

\subsubsection{MultiNest}

The mainstay of Bayesian posterior distribution evaluation has for many years been Monte Carlo methods, in particular Monte Carlo Markov Chain (MCMC) algorithms. Unfortunately it turns out that most MCMC incarnations generally have trouble accurately estimating the evidence term, in particular when the posterior has multiple modes. Indeed many of the various MCMC packages available do not directly calculate the evidence term. 

As noted before, the estimation of parameters does not require one to calculate the evidence term, but model selection does. A recent innovation in Monte Carlo methods is the MultiNest algorithm, which is specifically designed to accurately evaluate evidences \citep{Feroz08,Feroz09}.

\subsection{Application of Bayesian model selection to TTV search}

TTVs arise from the dynamics of a planetary system if there is more than one planet in the system. But as it is commonly known, the three body problem and higher is chaotic over long time scales. Such a signal could show up as a change in apparent orbital period, or the phase of the orbit.

Given these ambiguities we have then chosen to frame the search for a TTV signal as a Bayesian model selection problem. To every transit we observe we can unambiguously assign an epoch, given prior information on the orbital period, which in the case of GJ\,1214\,b is approximately 1.5804 days. We have calculated the probability of the linear no-TTV model versus models with TTV simulated with an N-body code.

\subsubsection{Non-TTV model}

Assuming no transit timing variation, we would expect that the relation between the time of mid-transit and the epoch is perfectly linear. In the Bayesian framework presented above, given the distribution of errors and the prior ranges, we can calculate the probability of this particular model itself and compare it directly to the probability of the model with a TTV signal.

As the prior ranges are included in the { overall} posterior probabilities of the models it is necessary to assign ranges to these priors, { i.e. the posterior probability takes the form of an average of the likelihood over the prior; see Eq.\,\ref{evidence}.}

The parameter $T_{0}$ in both the models is related to the definition of the epoch. It is simply the Julian date of the mid-transit of epoch 0. When the epoch is defined, this quantity is in principle known, but we cannot determine it exactly; hence, we introduce a systematic error. In a Bayesian framework we can take this into account by marginalising out $T_{0}$. The prior range of $T_{0}$ has been chosen to correspond to the largest error in the two measurements that pertain to epoch 0 in Table\,\ref{tab:timemin}.


\subsubsection{TTV models}

Unfortunately there is no known analytic solution to the general three body problem. Hence, to calculate the TTV arising to a third perturbing body in the system one has to rely on numerical orbital integration codes. One such code is {\sc Swift} \citep{swift}, which has been employed here. For our purposes we have adapted the {\sc FORTRAN} code used in \cite{Nesvorny2008}, with this adapted code we where able to satisfactorily reproduce results in that article.

For the purpose of this investigation it was assumed that GJ\,1214\,b is in a circular orbit, which is likely given the results in \cite{Charbonneau09} where the eccentricity is estimated to be less than 0.27. Further we assume that GJ\,1214\,b is perturbed by an planet in a coplanar circular orbit.

\begin{figure}
\centering
\includegraphics[width=\columnwidth]{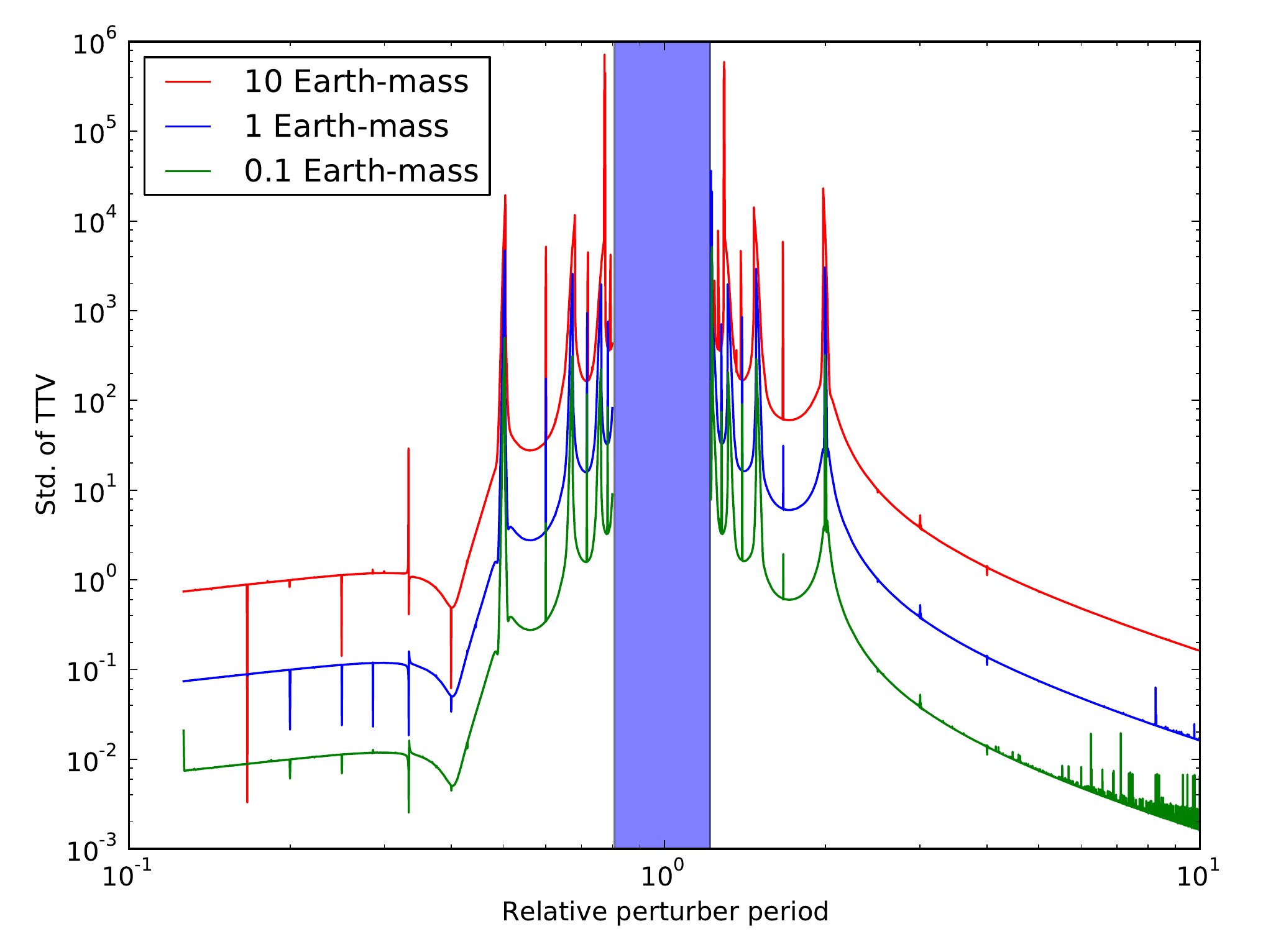}
\caption{Plot of the standard deviation, { in units of seconds}, of the TTV signal of GJ\,1214\,b calculated for different masses of the perturber. The various mean motion resonances are clearly visible. The blue shaded area marks perturber orbits that would make the orbit of GJ\,1214\,b unstable.}
\label{ttv-est}
\end{figure}

Given a trial mass $m$, period $P$ and mean longitude $\lambda$ of the hypothetical perturbing planet, and assuming the time of first transit to be $t=0$, the provided code calculates the time of all future transits taking into account the gravitational interaction between the planets. In the simple case of two coplanar circular orbits the argument of periapsis simply determines the angular separation of the two planets at the start of the integration. { The standard deviation of the TTV signal as a function of these parameters can be calculated numerically as done in Fig. \ref{ttv-est}.}

{ The majority of perturber parameter space will give rise to negligible TTV, but from Fig. \ref{ttv-est} two regions of perturber parameter space which can give rise to significant TTV, comparable to the uncertainty of our data, can be identified. One region consists of periods in the range from the 2:1 resonance, at a relative period of 0.5, to the inner edge of the instability strip. The other region is from the outer edge of the instability strip to the 2:1 resonance, at a relative period of 2. In both of these regions, perturbers down to a mass of about 0.1 Earth mass will give rise to TTV of a second or more. Hence we will use these ranges as prior ranges in the model selection. The two regions will be treated as two different models to be tested against each other. The upper mass range for these two models will be set by the accuracy of radial velocity data from \citet{Charbonneau09}, where radial velocity data with a accuracy of about 10m/s were presented. We estimate that in the outer model, perturbers heavier than 6 Earth masses are excluded and for the inner model, perturbers heavier than 5 Earth masses are excluded by RV data. The two TTV models are presented in Table \ref{tab:ttvmodels} together with the non-TTV (i.e. linear) model, and their priors. }

\begin{table*}
\caption{Table of parameters for the non-TTV and two TTV models that have been analysed.}
\label{tab:ttvmodels}
\centering
	\begin{tabular}{lccccccc}
	\hline\hline
	Parameter		   & Prior range & Prior type & Mean & 	St. Dev.	&	MAP & ln(evidence) & Probability \\
	\hline
	\multicolumn{6}{l}{Linear $T_{mid} = P\cdot E+ T_{0}$} & 134.98 & 0.9999\\
	\hline
	$T_{0}$\tablefootmark{a}& $\pm 0.0002$	& Uniform	&  2455320.535732&    & 2455320.535533  &&\\
	$P$	in days			    & $\pm 0.0000002$	    & Uniform	& 1.580405       & 0.000015  & 1.580340	&&\\
	\hline
	\multicolumn{6}{l}{Interior perturer: $T_{mid} = TTV(a,m,\lambda,E)+T_{0}$} & 122.09 & $3\cdot 10^{-6}$\\
	\hline
	$T_{0}$\tablefootmark{a}         &  $\pm 0.0002$	 & Uniform 	&-0.000003 & 0.0001 & 0.0001 &&\\
	$\lambda$				         &  $0-2\pi$  	     & Uniform  &3.1       & 1.2    &1.9       &&\\
	$P$ in days                      &  $0.76$--$1.23$    & Jeffrey  & 0.9      & 0.1    &0.9      &&\\
	$m$ in $M_{\sun}$\tablefootmark{b} &$0.0000003$--$0.0000135$  & Uniform & 0.0000009 &0.000001 &0.0000006    &&\\
	\hline
	\hline
	\multicolumn{6}{l}{Exterior perturber: $T_{mid} = TTV(a,m,\lambda,E)+T_{0}$} & 122.71 & $5\cdot 10^{-6}$\\
	\hline
	$T_{0}$\tablefootmark{a}      &  $\pm 0.0002$	   & Uniform  &0.000002 & 0.0001   & 0.00004&&\\
	$\lambda$				      &  $0-2\pi$  	       & Uniform  &3.1      & 1.4      & 6.0    &&\\
	$P$ in days                   &  $1.91$--$3.18$    & Jeffrey  & 2.7     & 0.3     & 2.7  &&\\
	$m$ in $M_{\sun}$\tablefootmark{b} & $0.0000003$--$0.000018$ & Uniform  & 0.0000002 & 0.000002   &0.0000005 &&\\
	\hline
	\end{tabular}
\tablefoot{\tablefoottext{a}{Nuisance parameter marginalised out.}\tablefoottext{b}{These are scale parameters and should have a Jeffrey prior according to \cite{gregory05}.}}
\end{table*}


\begin{figure*}
\centering
\begin{tabular}{ccc}
\includegraphics[width=0.29\linewidth]{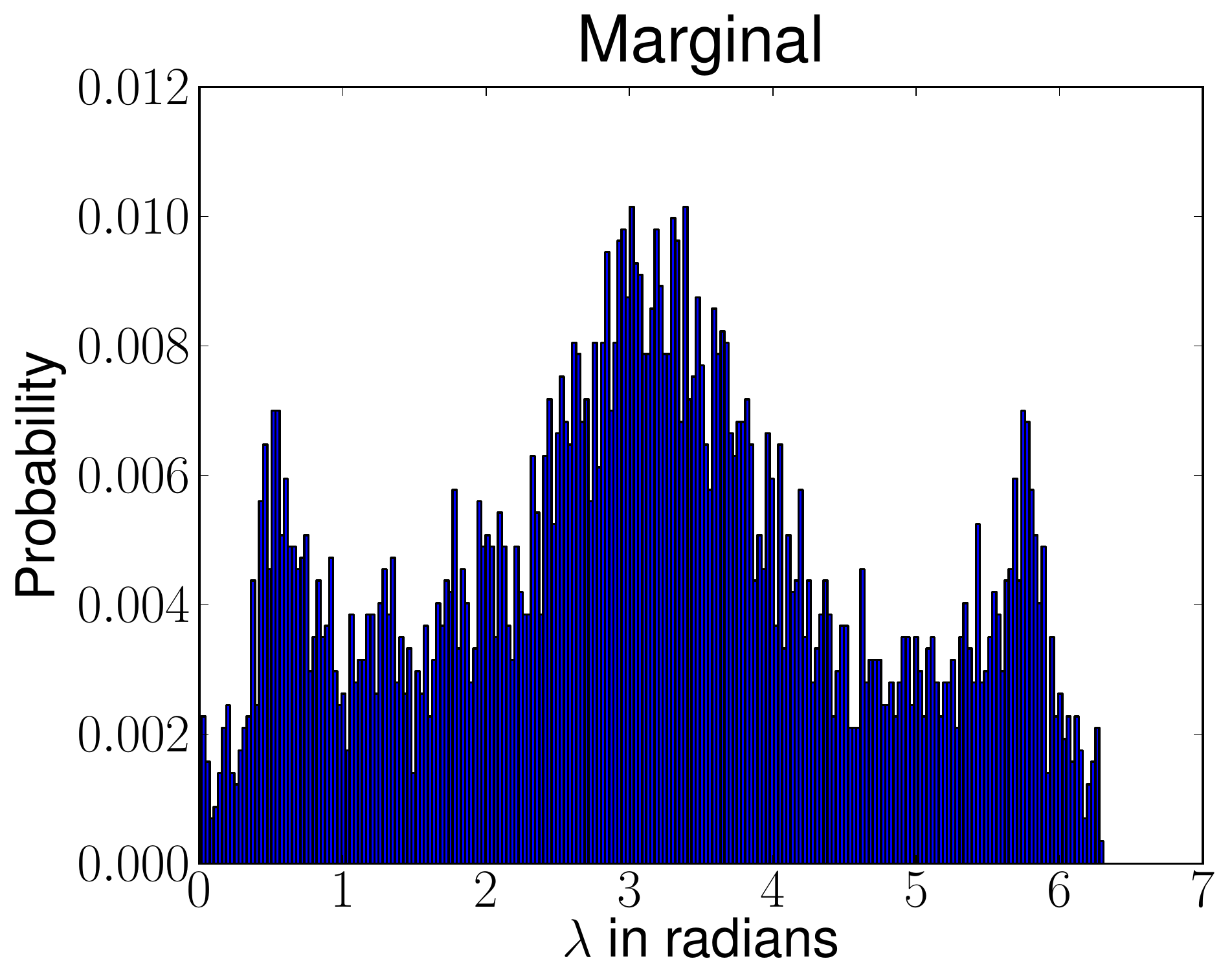} &
\includegraphics[width=0.29\linewidth]{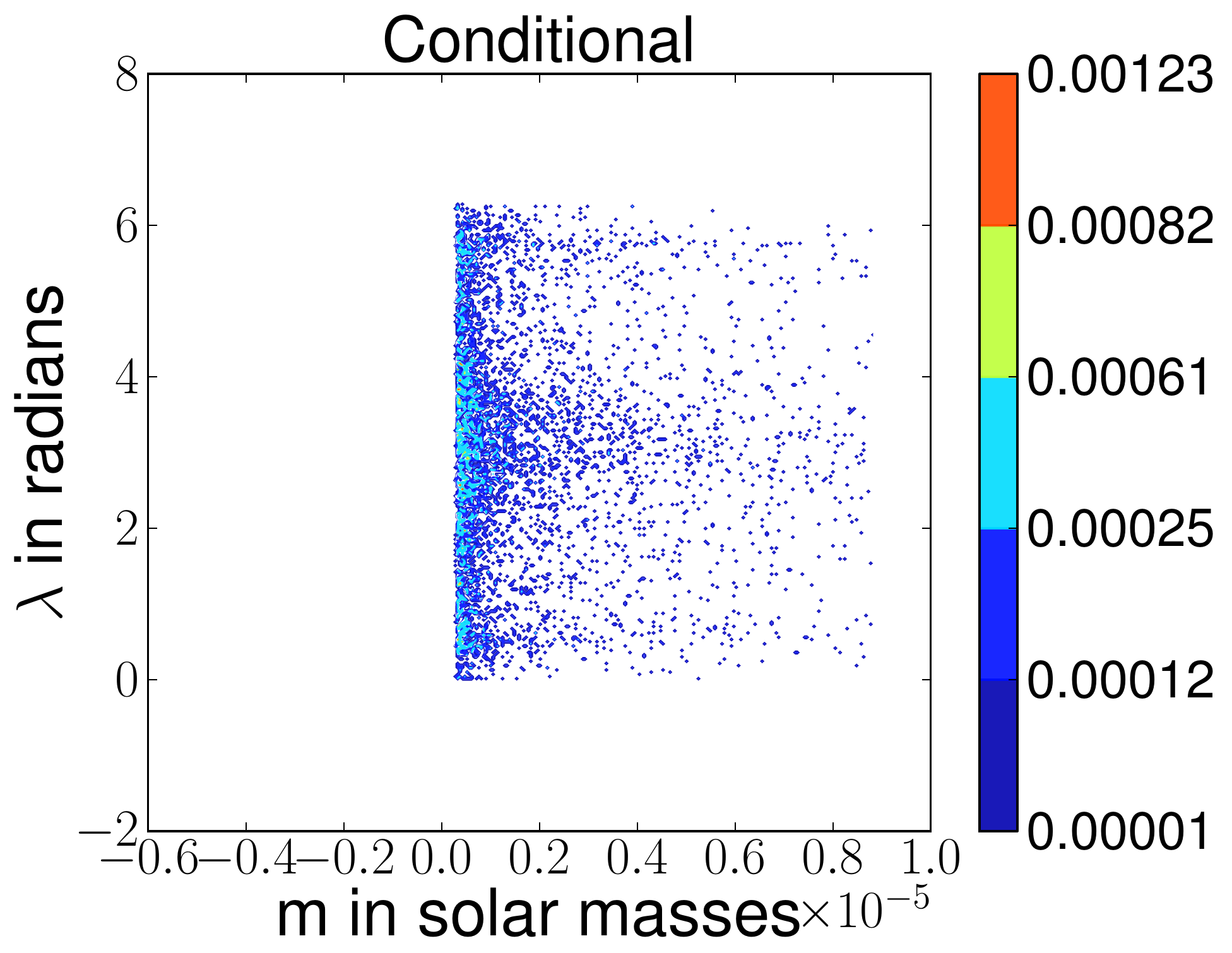} &
\includegraphics[width=0.29\linewidth]{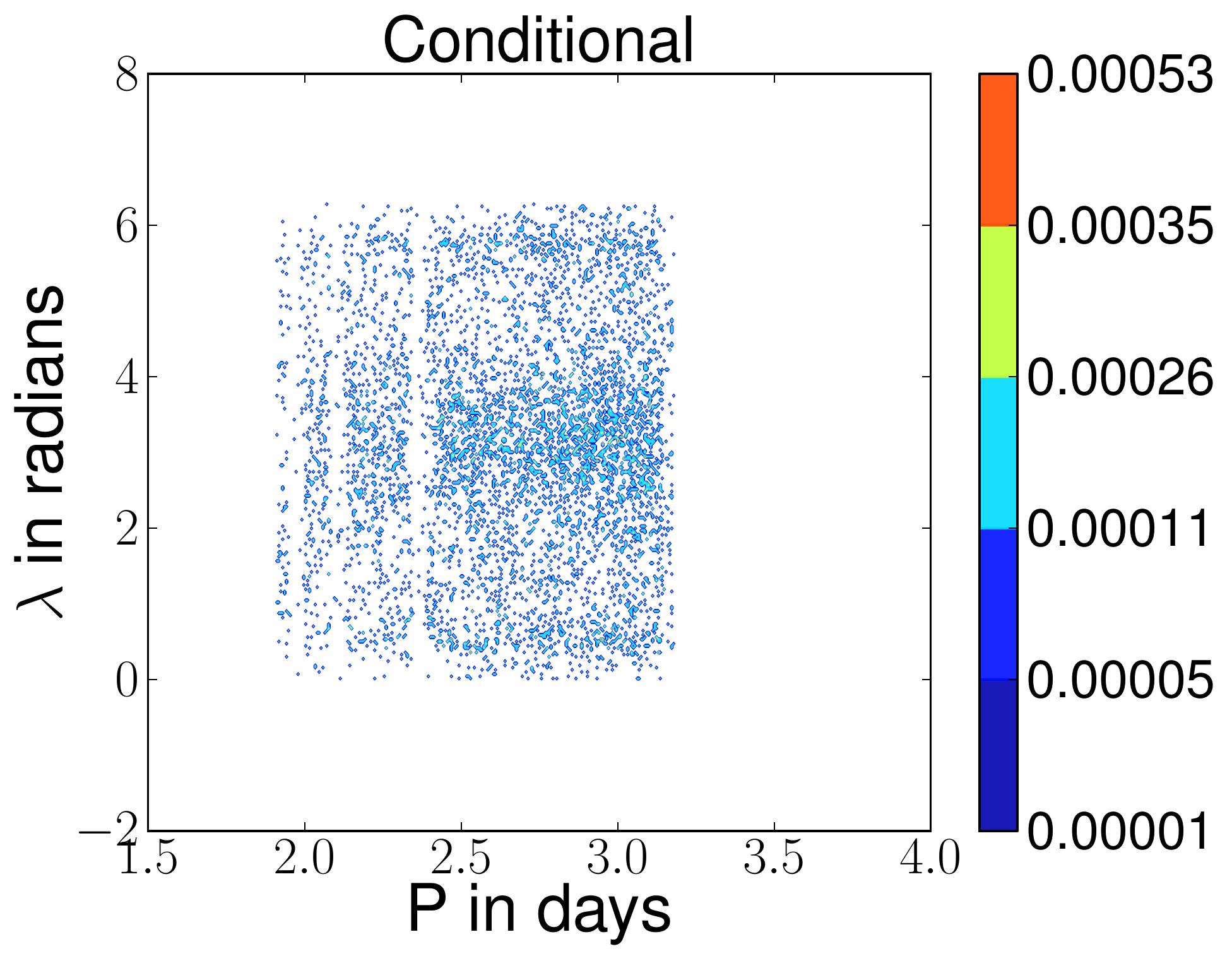} \\

&

\includegraphics[width=0.29\linewidth]{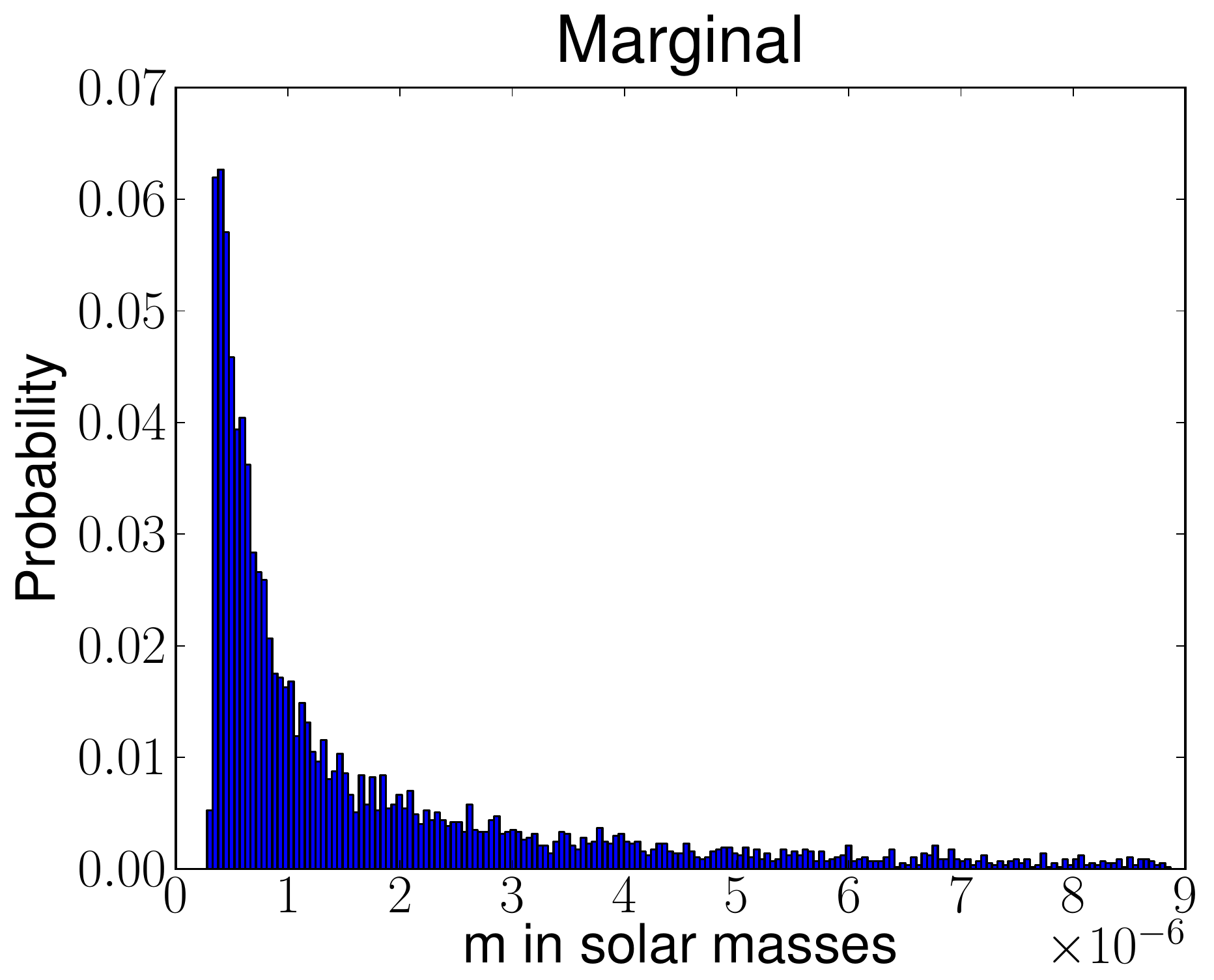} &
\includegraphics[width=0.29\linewidth]{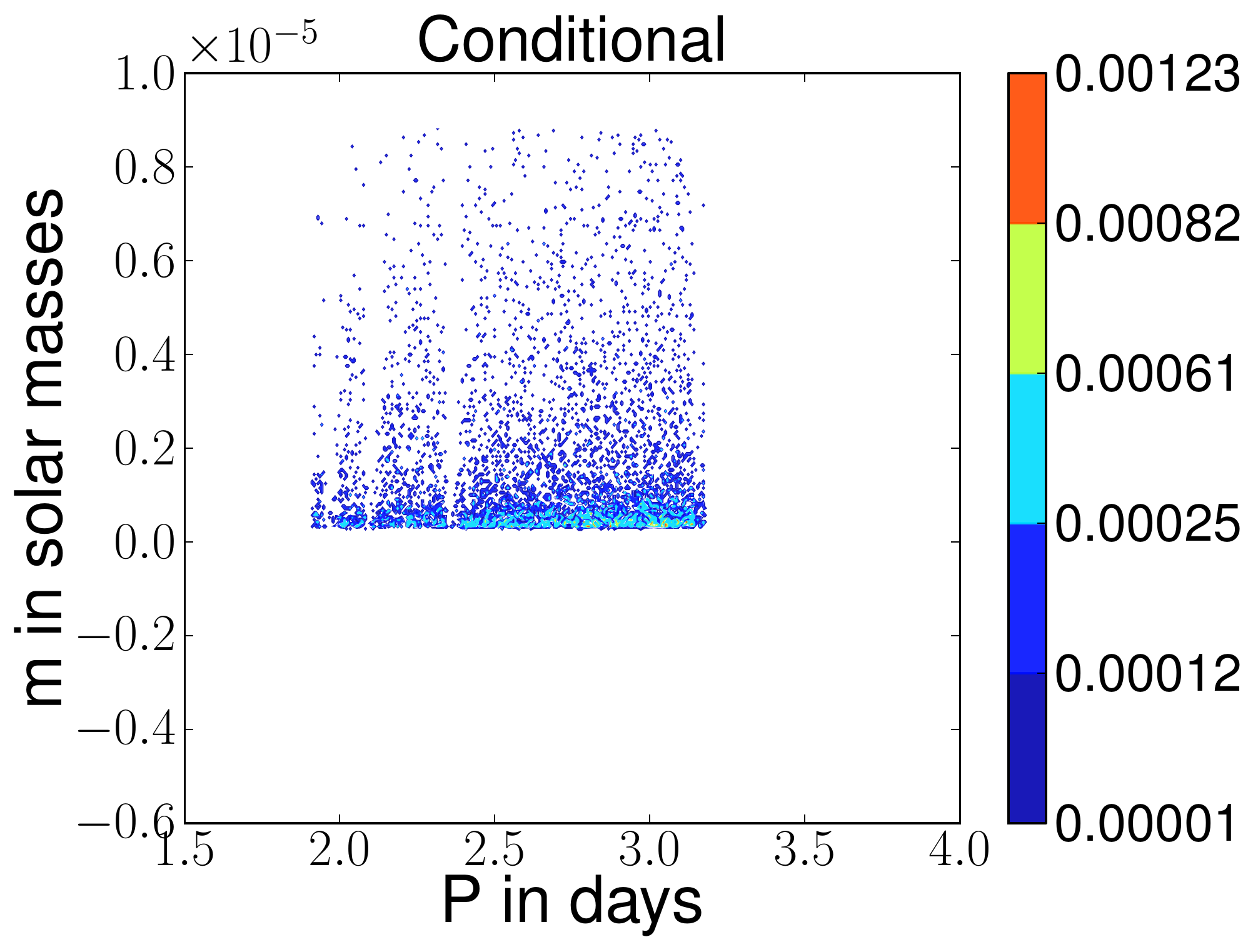} \\

&  &

\includegraphics[width=0.29\linewidth]{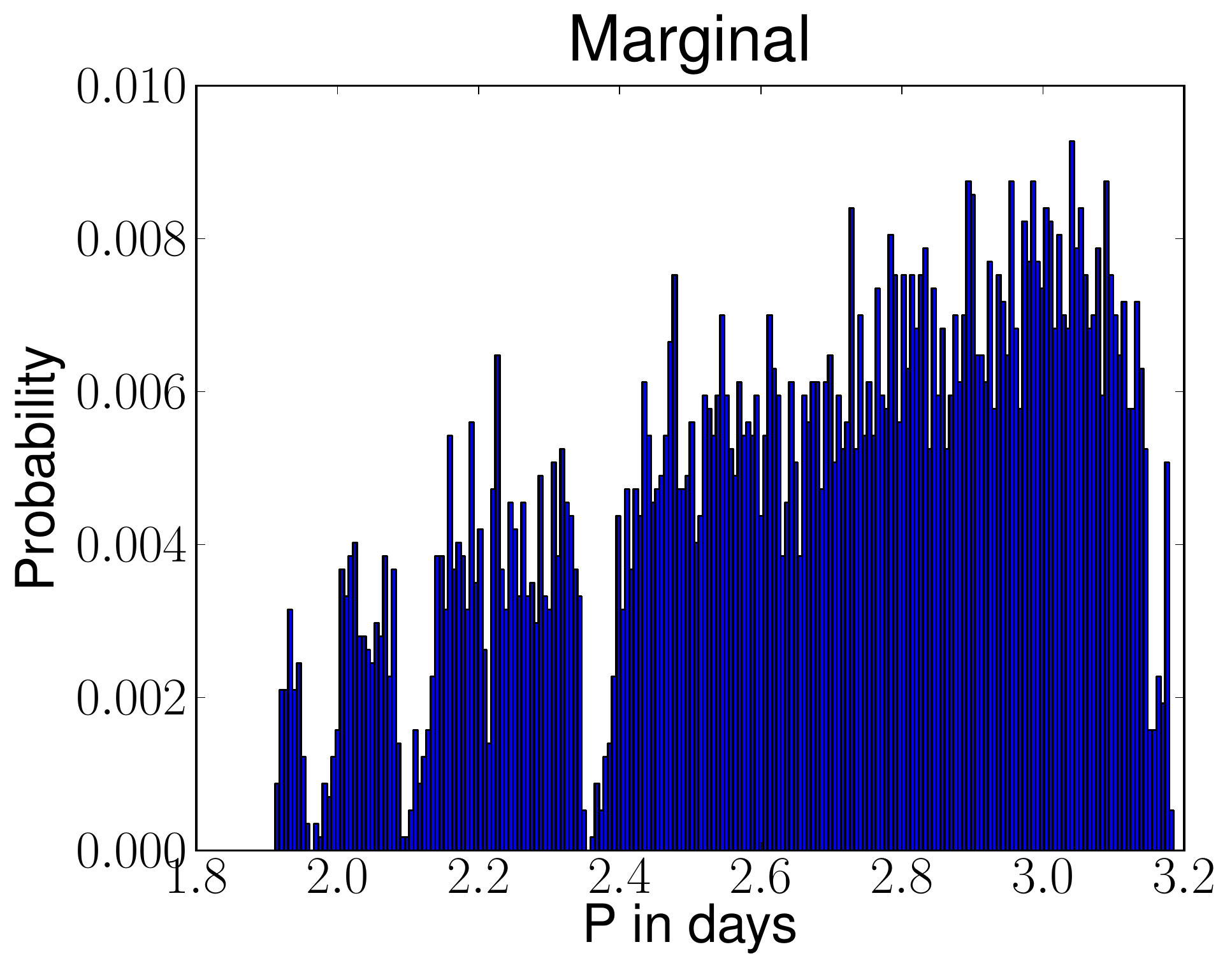} \\

\end{tabular}
\caption{Marginal probability and conditional probabilities for the TTV model assuming an exterior perturber from MultiNest. The parameter $T_0$ is a nuisance parameter, which is integrated out, hence the marginal and conditional probabilities have not been plotted.}
\label{marg-sin}
\end{figure*}

\begin{figure*}
\centering
\begin{tabular}{ccc}
\includegraphics[width=0.29\linewidth]{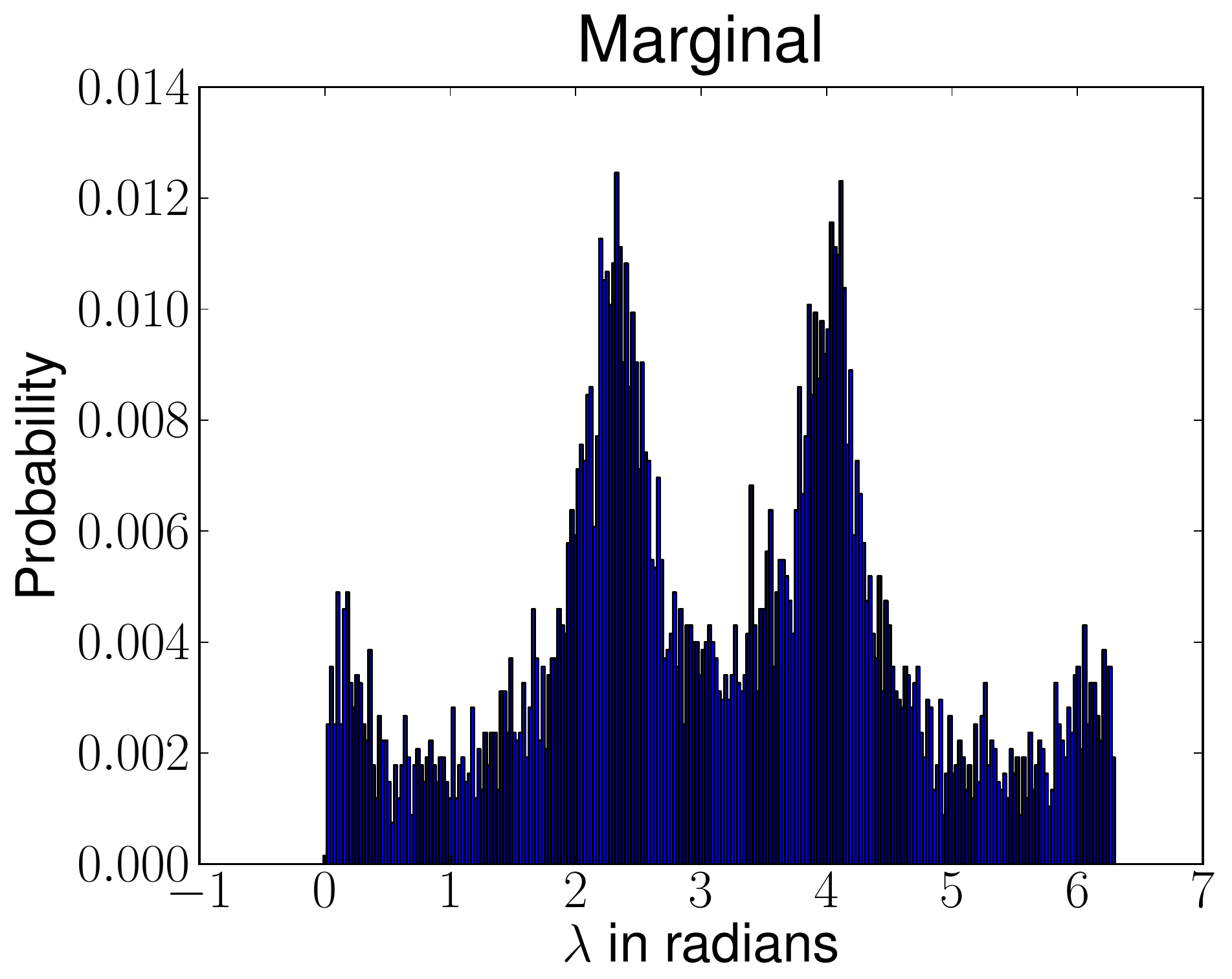} &
\includegraphics[width=0.29\linewidth]{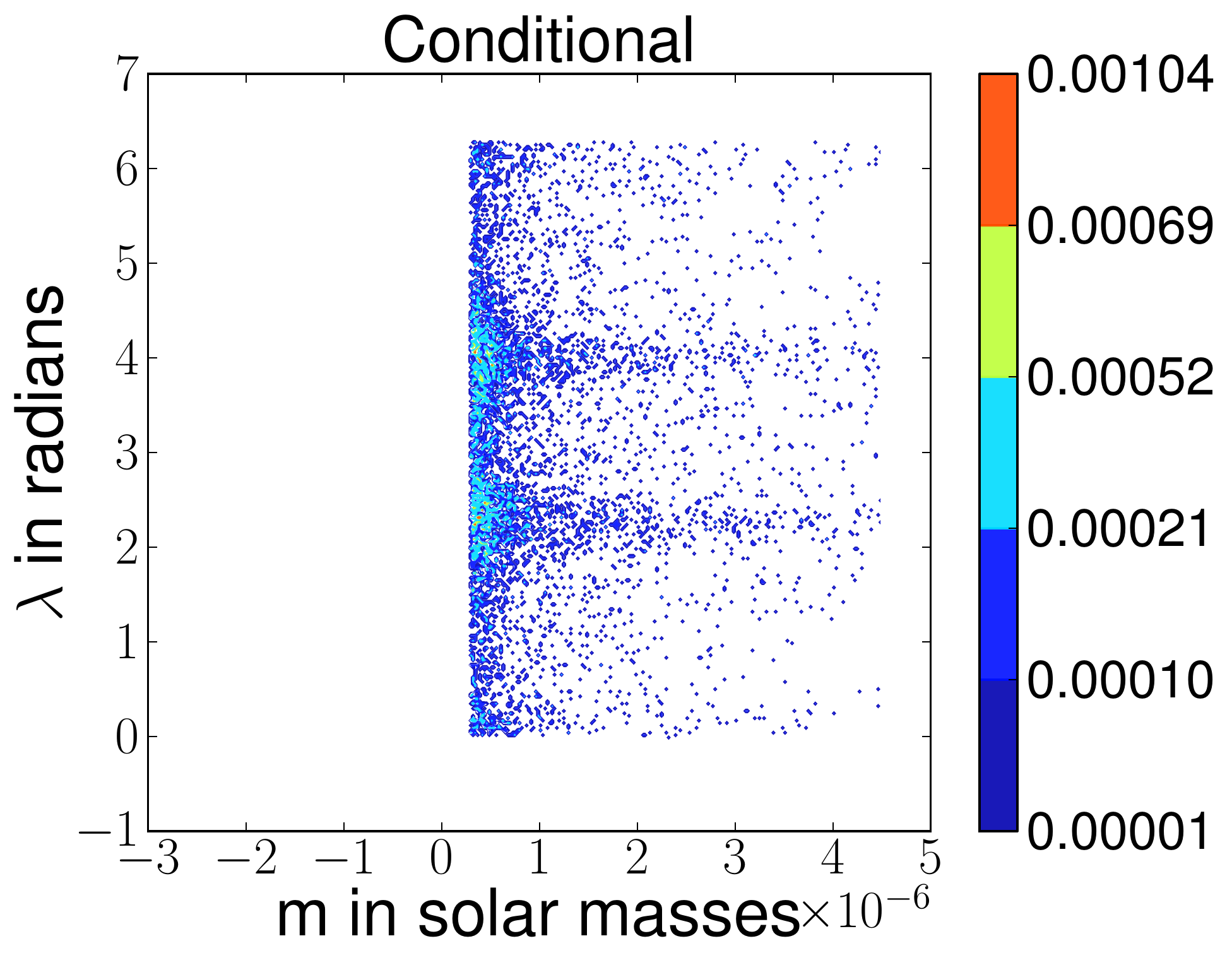} &
\includegraphics[width=0.29\linewidth]{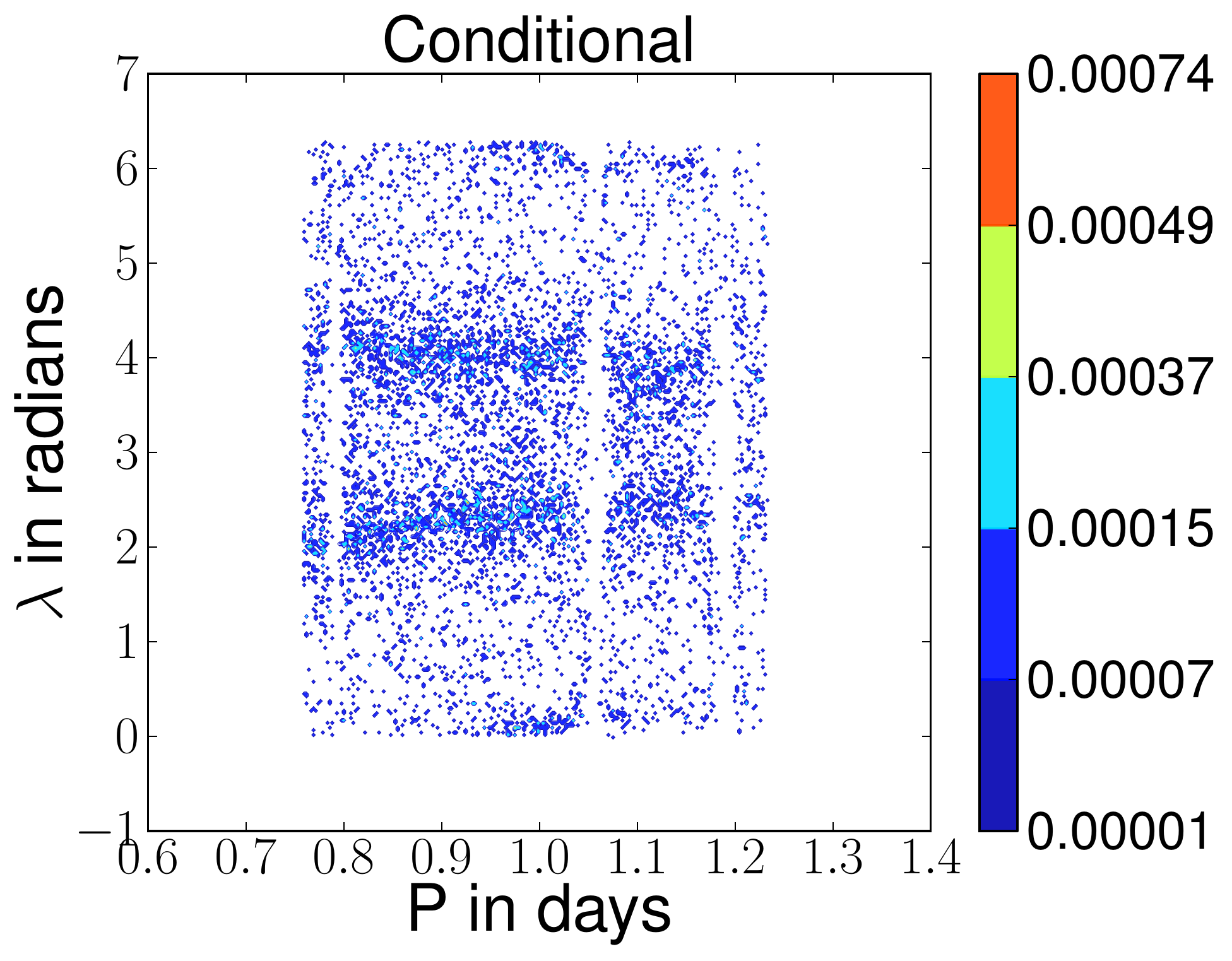} \\

&

\includegraphics[width=0.29\linewidth]{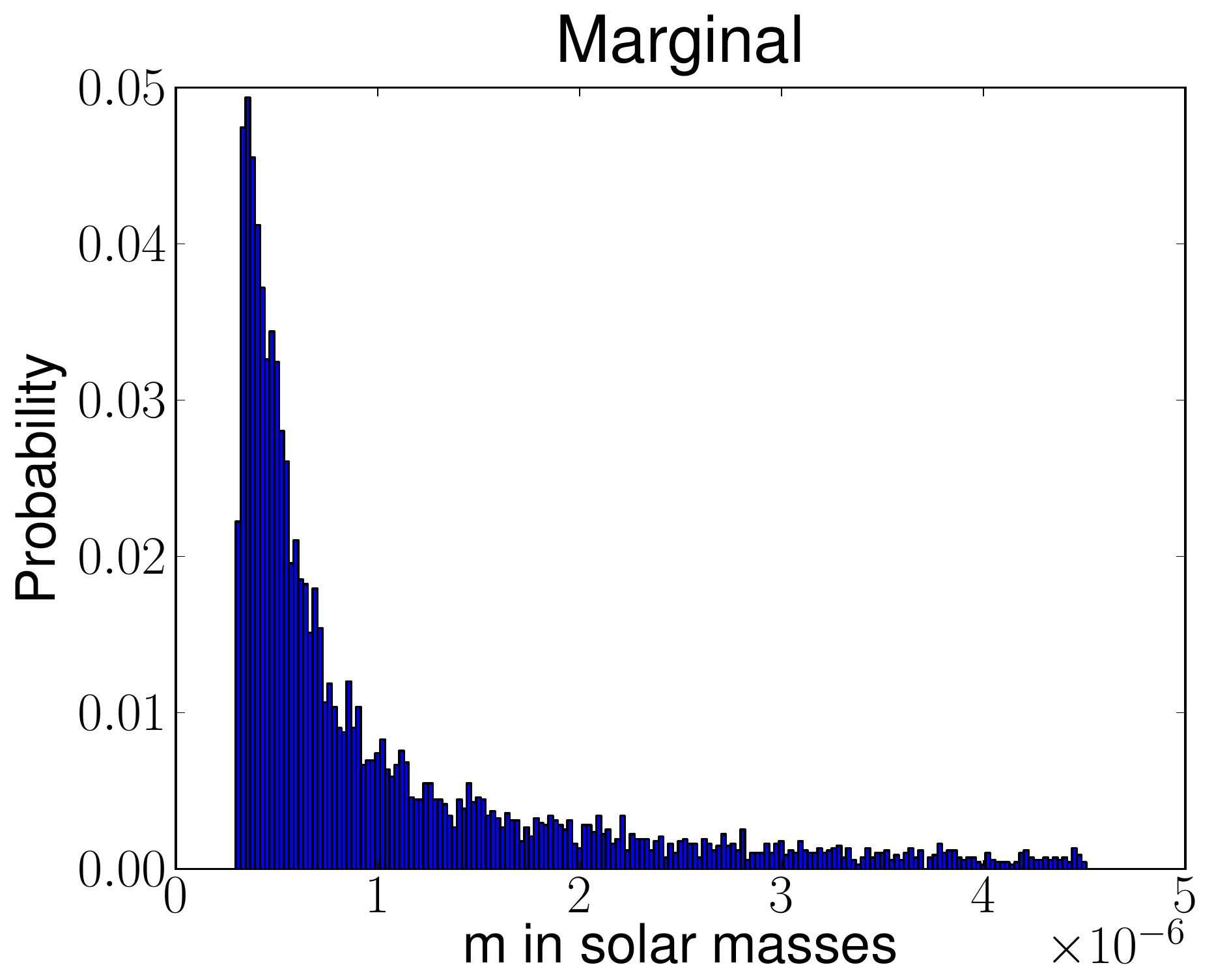} &
\includegraphics[width=0.29\linewidth]{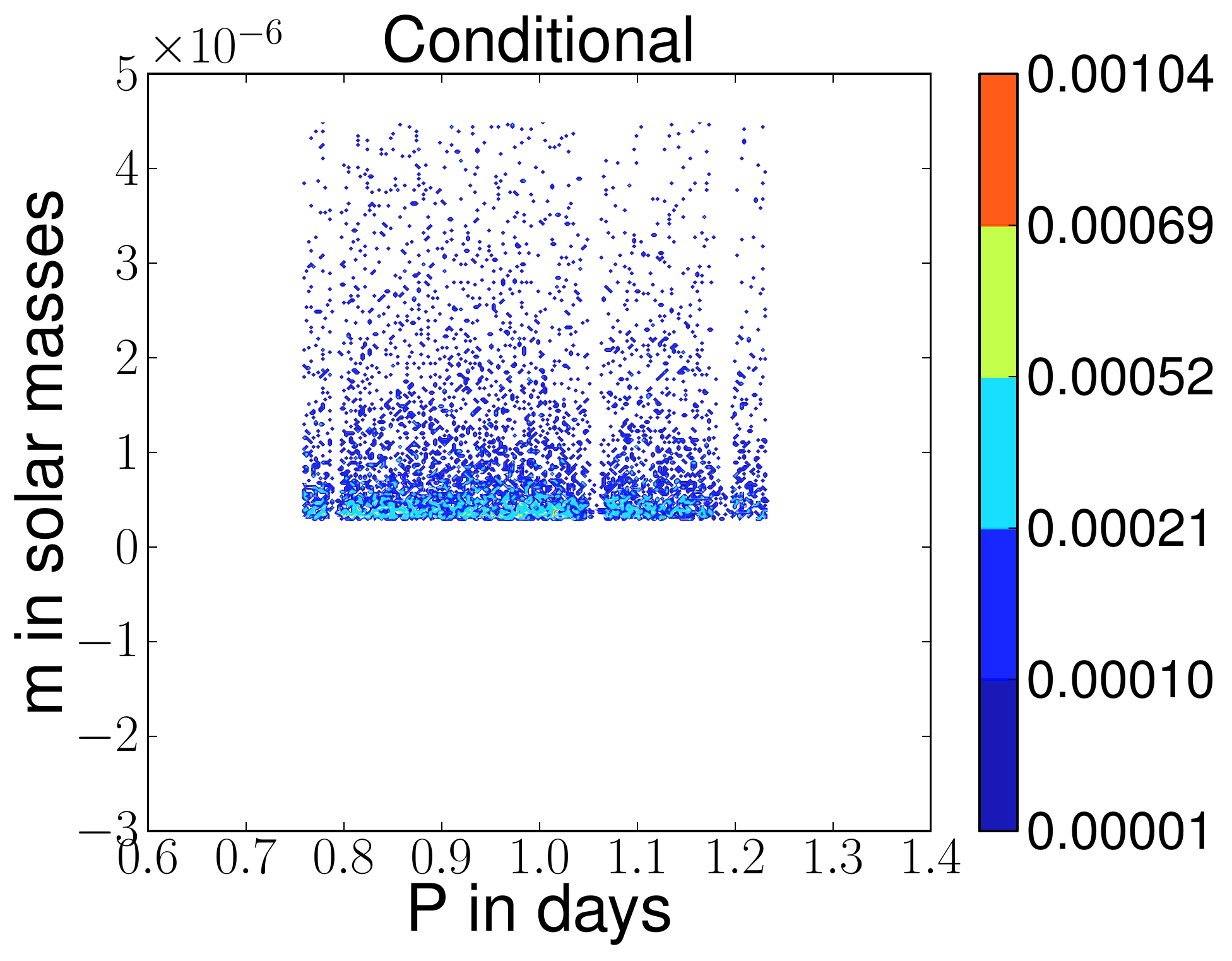} \\

&  &

\includegraphics[width=0.29\linewidth]{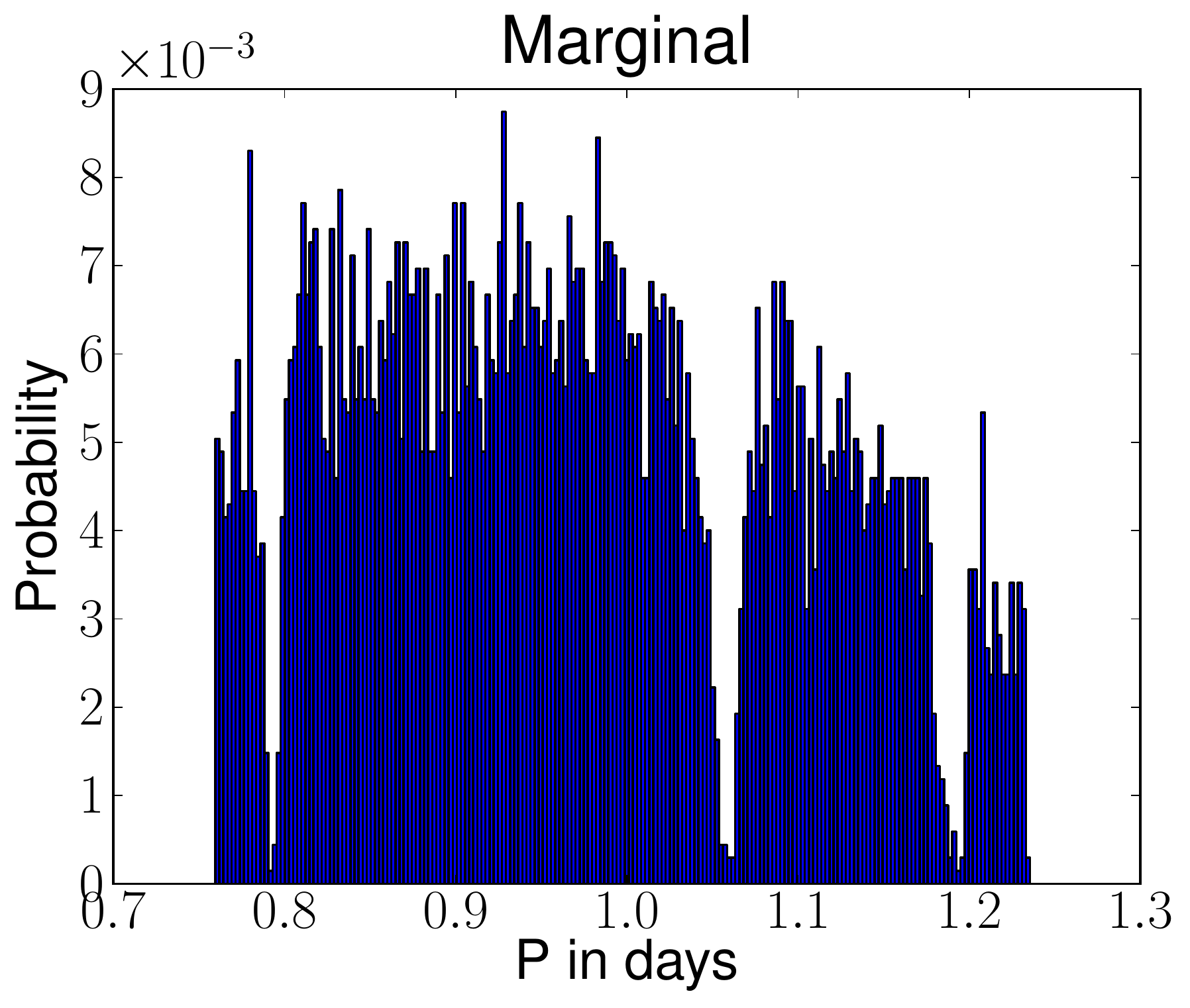} \\

\end{tabular}
\caption{Marginal probability and conditional probabilities for the TTV model assuming an interior perturber from MultiNest. The parameter $T_0$ is a nuisance parameter, which is integrated out, hence the marginal and conditional probabilities have not been plotted.}
\label{marg-sin2}
\end{figure*}


\subsection{Discussion}
The final results of these three models have been summarised in Table\,\ref{tab:ttvmodels}, along with the prior ranges of the parameters. We conclude that there is a very strong preference for the no-TTV model, given the calculated overall probabilities of the three models, because the relatively low quality, in the form of large uncertainties, of the available data does not allow us to reach complicated conclusions.
  
It should be noted that posterior probability of the model includes these prior ranges, hence the probability of the model can be thought of as the probability that a planet in this part of parameter space explains the transit times of GJ\,1214\,b. But note that the probability of a model is normalised by the prior volume, so a model with larger prior ranges or more parameters will be less probable, unless heavily supported by data. Large prior ranges are equivalent to lack of prior knowledge, hence one needs strong evidence to accept a complicated model in absence of prior knowledge.

MultiNest is in essence a Monte Carlo code and it does produce posterior samples, which can be analysed similarly to the samples produced by conventional Monte Carlo codes to produce estimates of the posterior probability of the parameters and their correlations. This has been done in Figs. \ref{marg-sin} and \ref{marg-sin2} for the two TTV cases, respectively.

{ The posterior density of the period $P$ in the no-TTV shows good agreement with the estimate produced in section \ref{sec:period}. To evaluate the evidence term we have marginalised over a range of offsets to take account for the uncertainly in the timing of the zero epoch.

Similarly the plot of the posterior probabilities in Fig. \ref{marg-sin} and \ref{marg-sin2} shows the marginal probabilities of the parameters of a perturbing planet, assuming that the data is explained by a perturbing planet within the prior ranges. These marginal probabilities show preference for a light planet (small $m$). The gaps in the marginal posterior of the period $P$ occurs at strong resonances that would give rise to TTV much larger than what is observed.}

\section{Conclusions}
In this paper we have presented and analysed 11 new transit light curves of the exoplanet GJ\,1214\,b. In addition new analyses of three previously published transits are presented. With these new data it has been possible to improve the previously estimated ephemeris and physical properties of GJ\,1214 and GJ\,1214\,b. The calculated physical properties are consistent with previously published values and the uncertainty in the orbital period has been reduced by a factor of approximately two -- $P = 1.58040456 \pm 1.6\cdot10^{-7}$.

Furthermore a stringent analysis, { incorporating available prior knowledge in the form of orbital mechanics and previous RV investigations, of whether transit timing variation can explain the data has been undertaken via Bayesian methods.} Specifically we have calculated the probability of three models for explaining the data, two with a TTV and one without, assuming {\it a priori} that one of the models is true. { This analysis has revealed that the models with TTVs are highly improbable compared to the simple model assuming no TTV. That is, the given data does not allow us to conclude that there is a planet in the mass range 0.1--5 Earth-masses and the period range 0.76--1.23 or 1.91--3.18 days. To be able to reach such conclusions we would need many more consistent data points and/or higher accuracy.}

{ A planet with a greenhouse warming and albedo similar to Earth at a period of approximately 4.5 days in the GJ\,1214 system, corresponding to a period relative to GJ\,1214\,b of 3, would have a surface temperature of about 80$^{\circ}$\,C. Conversely a planet in this orbit with an albedo similar to Venus and a greenhouse warming of 0 would have a surface temperature of about 0$^{\circ}$\,C. Hence such a planet could be at the inner or outer edge of the habitable zone of GJ\,1214\,b depending on the parameters -- for high albedo and/or low greenhouse warming the habitable zone overlaps with the region of parameter space that can reasonably be investigated with a transit timing accuracy on the order on 10\,s; see Fig. \ref{ttv-est}. Because of the orbital resonance, habitable planets down to Mars-mass could potentially be revealed in the GJ\,1214\,b system with the already achieved timing accuracy. To investigate the habitable zone for planets more similar to Earth, a much higher accuracy on the order of 0.01\,s is called for.} 




\begin{acknowledgements}
In addition to data from the Danish 1.54m telescope at ESO La Silla Observatory, this article is based on (1) observations collected at the MPG/ESO 2.2\,m Telescope located at
ESO La Silla, Chile. Operations of this telescope are jointly
performed by the Max Planck Gesellschaft and the European Southern
Observatory. GROND has been built by the high-energy group of MPE in
collaboration with the LSW Tautenburg and ESO, and is operating as a
PI-instrument at the MPG/ESO 2.2\,m telescope;
(2) data collected at the Centro Astron\'{o}mico Hispano Alem\'{a}n
(CAHA) at Calar Alto, Spain, operated jointly by the Max-Planck
Institut f\"{u}r Astronomie and the Instituto de Astrof\'{i}sica de
Andaluc\'{i}a (CSIC);
(3) observations obtained with the 1.52\,m Cassini telescope at the
Astronomical Observatory of Bologna in Loiano, Italy.

The authors would like to acknowledge Johannes Ohlert from the Astronomie Stiftung Trebur, Germany and Thomas Sauer from Internationale Amateursternwarte e. V., for graciously contributing datasets from the Michael Adrian Observatory and Observaorium Hakos, respectively, via the Exoplanet Transit Database \citep{etd}. 

J.\ Southworth acknowledges support from STFC in the form of an Advanced Fellowship.
J.\ Surdej, F.\ Finet, D.\ Ricci and O.\ Wertz acknowledge support from the Communaut\'e fran\c caise de Belgique - Actions de recherche concert\'ees - Acad\'emie
universitaire Wallonie-Europe.

Research by T.C. Hinse is carried out at the Korea Astronomy and Space
Science Institute (KASI) under the KRCF (Korea Research Council of
Fundamental Science and Technology) Young Scientist Research
Fellowship Program. T.C. Hinse acknowledges support from KASI registered
under grant number 2012-1-410-02.

The contributions from K. A. Alsubai, M. Dominik, M. Hundertmark and C. Liebig were made possible by NPRP grant NPRP-09-476-1-78 from the Qatar National Research Fund (a member of Qatar Foundation). The statements made herein are solely the responsibility of the authors.

\end{acknowledgements}

\bibliographystyle{aa}
\bibliography{litteratur}

\end{document}